\documentclass[11pt]{article}
\usepackage{amsfonts}
\usepackage{amsmath}
\usepackage{amssymb}
\usepackage{amsthm}
\usepackage{appendix}
\usepackage[usenames,dvipsnames]{color}
\usepackage{comment}
\usepackage{graphicx}
\usepackage{mathrsfs}
\usepackage{natbib}
\usepackage{showidx}
\usepackage{bm}
\usepackage{booktabs}
\usepackage{url}

\newcommand{\Pb}{\mathbb{P}}
\newcommand{\R}{\mathbb{R}}
\newcommand{\Sy}{\mathbb{S}}

\newcommand{\Cs}{\mathcal{C}}
\newcommand{\Fs}{\mathcal{F}}
\newcommand{\Ls}{\mathcal{L}}

\newcommand{\Ber}{\mathrm{Bernoulli}}

\newcommand{\Bin}{\mathrm{Binomial}}

\newcommand{\IGa}{\mathrm{Inv{\textrm -}Gamma}}

\newcommand{\Nor}{\mathrm{Normal}}
\newcommand{\Uni}{\mathrm{Uniform}}

\newcommand{\E}{\mathrm{E}}
\newcommand{\Corr}{\mathrm{Corr}}
\newcommand{\Cov}{\mathrm{Cov}}
\newcommand{\Var}{\mathrm{Var}}

\newcommand{\bc}{\boldsymbol{c}}
\newcommand{\be}{\boldsymbol{e}}

\newcommand{\bp}{\boldsymbol{p}}

\newcommand{\by}{\boldsymbol{y}}
\newcommand{\bz}{\boldsymbol{z}}

\newcommand{\bmu}{\boldsymbol{\mu}}
\newcommand{\bsigma}{\boldsymbol{\sigma}}

\newcommand{\brho}{\boldsymbol{\rho}}

\newcommand{\btheta}{\boldsymbol{\theta}}
\newcommand{\bxi}{\boldsymbol{\xi}}

\newcommand{\bSigma}{\boldsymbol{\Sigma}}
\newcommand{\bTheta}{\boldsymbol{\Theta}}
\newcommand{\bXi}{\boldsymbol{\Xi}}

\newcommand{\bOne}{\mathbf{1}}

\newcommand{\mC}{\mathbf{C}}
\newcommand{\mD}{\mathbf{D}}

\newcommand{\mI}{\mathbf{I}}
\newcommand{\mJ}{\mathbf{J}}
\newcommand{\mS}{\mathbf{S}}

\newcommand{\diag}{\mathrm{diag}}
\newcommand{\diff}{\mathrm{d}}

\newcommand{\iid}{\mathrm{iid.}}
\newcommand{\ind}{\mathrm{ind.}}
\newcommand{\logit}{\mathrm{logit}}

\newcommand{\sss}[1]{\scriptscriptstyle{#1}}

\newcommand{\red}[1]{\textcolor{red}{#1}}

\newtheorem{proposition}{Proposition}
\newtheorem{result}{Result}

\begin{document}

\title{Using Joint Random Partition Models for Flexible Change Point Analysis in
Multivariate Processes}
\author{Jos\'e J. Quinlan \\ Department of Statistics, Pontificia Universidad Cat\'olica de Chile, Chile \\
jjquinla@uc.cl \and Garritt L. Page \\ Department of Statistics, Brigham
Young University, USA \\ page@stat.byu.edu \and Luis M. Castro \\
Department of Statistics, Pontificia Universidad Cat\'olica de Chile, Chile \\
Millennium Nucleus Center for the Discovery of Structures in Complex Data, Chile\\
Centro de Riesgos y Seguros UC, Pontificia Universidad Católica de, Chile, Chile\\ mcastro@mat.uc.cl}
\maketitle

\begin{abstract}
    Change point analyses are concerned with identifying positions of an ordered stochastic process that undergo abrupt local changes of some underlying distribution. When multiple processes are observed, it is often the case that information regarding the change point positions is 
    shared across the different processes. This work describes a method that
    takes advantage of this type of information. Since the number and position of change points can be described through a partition with contiguous clusters, our approach develops a joint model for these
    types of partitions. We describe computational strategies associated with our approach and illustrate improved performance in detecting change points through a small simulation study. We then apply our method to a financial data set of emerging markets in Latin America and highlight interesting insights discovered due to the correlation between change point locations among these economies. 
\end{abstract}

\section{Introduction}\label{Introduction}

Change point analyses identify times or positions of an ordered stochastic process that undergo abrupt local changes. These abrupt changes are typically seen as  shifts in expectation, variability, or shape of an underlying distribution (or some combination of the three). Methods that detect change points have been employed in a variety of fields, including finance \citep{2021}, climatology 
\citep{Gupta_etal:2021}, and ecology \citep{jones_etal:2021} to name a few. Due to this, many change point methods have been proposed in the statistical literature both in a univariate \citep[see for example][]{ArellanoValleEtAl2013} and a multivariate setting \citep[see][for a comprehensive review]{truong:2020}.

\begin{figure}[ht!]
    \begin{center}
        \includegraphics[scale=0.75]{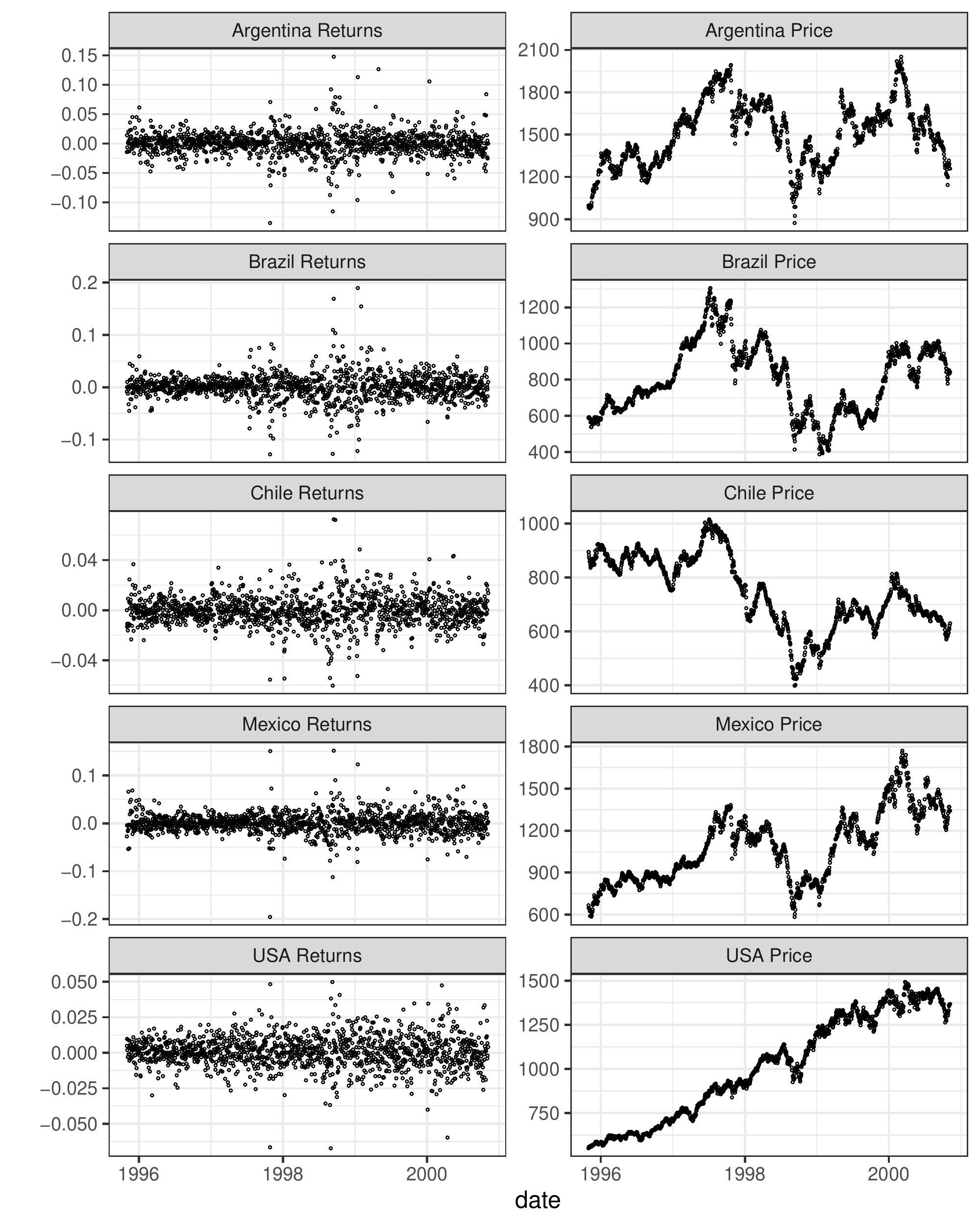}
        \caption{Daily returns and price of stock market indexes of Argentina, Brazil, Chile, Mexico and USA. Returns are calculated using $R_{t}=(P_{t}-P_{t-1})/P_{t-1}$ for $t\in\{2,\ldots,n\}$,
        where $P_{t}$ is the price of day $t$.}
    \end{center}
    \label{fig:data}
\end{figure}

The phenomenon that motivates our research is the so-called {\it financial contagion} or simply {\it contagion}. This phenomenon can be understood as the spread of financial crises from one country to another \citep[see for example][among others]{MR-962,Valdes2000,FilletiEtAl2008}. To illustrate {\it contagion} consider the price and returns of the five markets displayed in Figure \ref{fig:data}. Note that the overall trend of the price in the Latin American markets (Argentina, Brazil, Chile, and Mexico) seem to coincide. However, the USA market (Dow Jones index) presents a different trend during the same observation period (1995 to 2001). It is important to note that, in the second half of 1998, Dow Jones suffered a slight crash (a change in volatility according to Figure \ref{fig:data}, left column) due to the Russian financial crisis and the Long Term Capital Management episode. Consequently, we hypothesize that a change point in a mature market such as the US could produce change points in emerging markets such as those from Latin American, or simply, the {\it financial contagion} between the US market and Latin American markets could increase the chance of change points occurring in the later markets when they occur in the former. Consequently, the method we develop will incorporate dependence between change point probabilities across multiple processes which could potentially improve the ability of detecting a change point compared to an independent model.

One commonly used approach in the statistical literature for detecting time-series change points is based on product partition models (PPM). These models, which were introduced by \cite{Barry1992}, assume that (a) the number and positions of change points are random and, (b) observations within the same block are assumed to follow the same distribution. Thus, the inferential problem reduces to identifying a partition where each cluster is a collection of consecutive data points and then estimate parameters associated with each cluster's assumed data model. From a Bayesian viewpoint, a prior distribution on the space of partitions, which are restricted to be contiguous, is needed. \cite{Barry1992} use a prior for which prior change point probabilities are based on the so-called cohesion function studied in \cite{Yao1984}. Since \cite{Barry1992} many other PPM type approaches to change point analysis have be developed \cite[see, for example][]{Loschi2002,Loschi2003,Loschi2005,LoschiCruz2005,Loschi2010,Martinez2014,garcia2019,pedroso2021multipartition}.

Most of the existing approaches based on PPMs for detecting change points in time series treat the series individually. However, in the presence of {\it contagion}, the information available from several series could improve the accuracy of the change point detection mechanism compared to when series are treated individually. In general, the strategies for detecting change points in the multivariate context focus on detecting changes in the joint distribution of the coordinates of a multivariate process across time.
For example, \cite{Cheon2010} developed a Bayesian model for detecting changes in the mean and variance when data follow a multivariate normal distribution. Their approach considers a latent vector that identifies change point positions for partitioning the observations. 
\cite{Nyamundanda2015} proposed a Bayesian latent variable PPM, the so-called product partition latent variable model (PPLVM), providing a flexible framework for detecting multiple change point detection in multivariate data. The key feature of the PPLVM is that it can be used to detect distributional changes in the mean and covariance of the series, even in high-dimensional settings. Recently, \cite{jin2021} developed a hierarchical Bayesian model for changes in the mean process were the prior distribution for change point configurations arises from a Poisson-Dirichlet process, restricted to contiguous partitions \citep[see][for more details]{Martinez2014}.



In contrast to the approaches described above, there are models that do not introduce explicitly a distribution for partitions of contiguous clusters. For example, \cite{737745} considered a procedure for detecting change points by minimizing a cost function, which detects the optimal number and location of change points with a linear computational complexity under mild conditions. \cite{tveten2021scalable} also minimize a cost function when searching for change points in cross-correlated processes. \cite{matteson2014} proposed a robust nonparametric method using a divergence measure based on Euclidean distances. With this method, the authors showed that it is possible to detect any distributional change within an independent sequence of random variables without making any distributional assumption beyond the existence of the $\alpha$-th absolute moment, for some $\alpha\in(0,2)$. \cite{EJS1809} proposed a novel change point detection algorithm based on the Kolmogorov–Smirnov statistic and showed that it is nearly minimax rate optimal under suitable conditions. 

Other approaches for detecting changes in a multivariate process that is more inline with what we proposed assume that each process has its own change point structure. \cite{harle2016} introduce a set of independent binary vectors whose entries indicate which coordinate of the multivariate process changes. They do this using a composite marginal likelihood based on Wilcoxon's rank-sum test and a suitable prior for the binary vectors. Their approach allows us to include dependence among change points between different coordinates. Moreover, \cite{Fan:2017} introduce a set of change point indicator variables for each coordinate and time such that the prior change point probabilities for all coordinates at a fixed time are the same but change through time.

Our approach, which is motivated by the {\it contagion} phenomena, considers simultaneous changes in all parameters associated with a particular process dealing with multiple univariate time series (not all of which are necessarily the same type of response). Thus, each series has its change point structure where the change points between them may or may not coincide in time.  We base our approach on elements from the method described in \cite{Fan:2017} combined with a novel multivariate extension of the PPM approach of \cite{Barry1992}. The resulting method takes advantage of the existing correlation in change point locations between series. This strategy requires specifying a joint prior distribution for a collection of partitions. Constructing these types of dependent partition models over a series of partitions has only very recently been considered in the literature by, for instance, \cite{carlosetal,page2021}. The method we present is the first work that we are aware of that considers jointly modeling contiguous partitions.

The remainder of the article is organized as follows. Section \ref{Preliminaries} provides notation and background to the change point PPM. In Section \ref{JointPriorDistribution} we describe our approach of incorporating dependence between change point probabilities, provide some theoretical properties and details with regards to computation. Section \ref{simulation} describes a numerical experiment designed to study our method's ability to detect change points, and in Section \ref{sec:application} we apply our approach to the finance data concerning stock market returns of five countries. We close the paper with some concluding remarks in Section \ref{discussion}.  

\section{Background and Preliminaries}\label{Preliminaries}


To make the paper self-contained, we start this section with a background related to PPMs, introducing some notation we will use throughout the paper.

\subsection{Partition Definition and Notation}\label{PartitionDef}

Without loss of generality, consider $L>1$ time series $\by_{i}=(y_{i,1},\ldots,y_{i,n})^\top$, each of length $n>2$. Change points occur when the behavior of $\by_{i}$ undergoes sudden changes at unknown times. These times of sudden changes partition $\{1,\ldots,n\}$ into $k_{i}$ contiguous sets, say $\rho_{i}=\{S_{i,1},\ldots,S_{i,k_{i}}\}$, for some $k_{i}\in\{1,\ldots,n\}$. Here, $S_{i,j}$ is the $j$th block and $k_{i}$ is the number of blocks in $\rho_i$. 
The space of these types of partitions will be denoted by $\Cs_{n}$.

An alternative, and commonly used, way to denote a partition $\rho_{i}$ is through a collection of $n$ cluster labels $\be_{i}=(e_{i,1},\ldots,e_{i,n})^\top$, such that $e_{i,t}=j$ if $t\in S_{i,j}$. Due to the contiguous structure of the $\rho_{i}$ that we consider, the cluster labels that define $\rho_{i}$ necessarily are such that $1=e_{i,1}\leq e_{i,2}\leq\cdots\leq e_{i,n}=k_{i}$ and $e_{i,t}-e_{i,t-1}\in\{0,1\}$, for all $t\in\{2,\ldots,n\}$. Therefore, change point locations can be identified using $\be_i$. Letting $\tau_{i,0}=0$ and $\tau_{i,j}=\max\{t\in\{1,\ldots,n\}:e_{i,t}=j\}$ for all $j\in\{1,\ldots,k_{i}\}$, we have that $S_{i,j}=\{\tau_{i,j-1}+1,\ldots,\tau_{i,j}\}$ and the set $\bm{\tau}_i = \{\tau_{i,1}+1,\ldots,\tau_{i,k_{i}-1}+1\}$ identifies the locations at which change points in $\by_{i}$ occur.

In a change point setting there is another way to represent $\rho_{i}$ that will prove to be very useful for our approach. It is based on a set of change point indicators $\bc_{i}=(c_{i,1},\ldots,c_{i,n-1})^\top$, such that $c_{i,t}=1$ if time $t+1$ is a change point in $\by_{i}$, and $c_{i,t}=0$ otherwise. Note that $\be_{i}$ and $\bc_{i}$ have a one-to-one correspondence through the relation $c_{i,t}=e_{i,t+1}-e_{i,t}$, for all $t\in\{1,\ldots,n-1\}$, where $e_{i,1}=1$ by construction. The number of change points can be identified using $\bc_i$ by noticing that $k_{i}=1+\sum_{t=1}^{n-1}c_{i,t}$. In what follows, we will use $\rho_{i}$, $\be_{i}$ and $\bc_{i}$ interchangeably. With the necessary notation introduced, we next describe the change point PPM. 

\subsection{Change Point Product Partition Models}\label{changepointPPM}

For the $i$-th sequence, the PPM is a discrete distribution on space $\Cs_{n}$ such that
\begin{equation}
    \Pb(\rho_{i}=\{S_{i,1},\ldots, S_{i,k_{i}}\})=
    \frac{\prod_{j=1}^{k_{i}}c(S_{i,j})}{\sum_{\rho_{i}\in\Cs_{n}}\prod_{j=1}^{k_{i}}c(S_{i,j})},
    \nonumber
\end{equation}
where $c(S_{i,j})$ is referred to as a cohesion function and measures the {\it a priori} belief that elements in $S_{i,j}$ co-cluster. The change point PPM as described in \cite{Barry1992}, \cite{Loschi2003}, and others, uses \cite{Yao1984}'s cohesion function to assign probabilities to each element in $\Cs_{n}$. \cite{Yao1984}'s cohesion function applied to contiguous $S_{i,j}$ results in
\begin{equation} 
    c(S_{i,j}\,;\,p_{i})=
    \left\{
    \begin{array}{lc}
         p_{i}(1-p_{i})^{\tau_{i,j}-\tau_{i,j-1}-1}, & \mathrm{if}\quad\tau_{i,j}<n  \\
         (1-p_{i})^{\tau_{i,j}-\tau_{i,j-1}-1}, & \mathrm{if}\quad\tau_{i,j}=n
    \end{array}
  \right.,
  \label{cohesion.Yao}
\end{equation}
for some $p_{i}\in[0,1]$ such that $p_{i}=\Pb(c_{i,t}=1\mid p_{i})$. Based on this cohesion we have that
\begin{equation}
    \sum_{\rho_{i}\in\Cs_{n}}\prod_{j=1}^{k_{i}}c(S_{i,j}\,;\,p_{i})=1.
    \nonumber
\end{equation}
 Thus, the change point PPM takes on the following form
\begin{align}
    \Pb(\rho_{i}=\{S_{i,1},\ldots,S_{i,k_{i}}\}\mid
    p_{i})=\prod_{j=1}^{k_{i}}c(S_{i,j}\,;\,p_{i})=p_{i}^{k_{i}-1}(1-p_{i})^{n-k_{i}}.
    \nonumber
\end{align}

Once the partition model has been specified, the key idea behind change point modeling from a partition perspective is that observations within the same block are assumed to follow a common distribution, whereas different distributions are assumed between blocks. Following \cite{Barry1992}'s approach, given $\rho_{i}$, the joint density of $\by_{i}$ is written as a product of $k_{i}$ {\it data factors}, also known as marginal likelihoods, which measure the similarity of observations within each block. More precisely,
\begin{align}
    \begin{split}
        f(\by_{i}\mid\rho_{i},\bxi_{i})&=
        \prod_{j=1}^{k_{i}}\Fs_{i}(\by_{i,j}\mid\bxi_{i}),
        \\
        \Fs_{i}(\by_{i,j}\mid\bxi_{i})&=
        \int_{\bTheta_{i}}\Ls_{i}(\by_{i,j}\mid\btheta_{i},\bxi_{i})
        \diff G_{i}(\btheta_{i}\mid\bxi_{i}),
    \end{split}
    \label{Data.distribution}
\end{align}
where $\by_{i,j}=(y_{i,t}:t\in S_{i,j})^\top$ and
$\Ls_{i}(\,\,\cdot\mid\btheta_{i},\bxi_{i})$ is a likelihood function indexed by the 
set of parameters $\btheta_{i}\in\bTheta_{i}$ which are block-specific, and $\bxi_{i}\in\bXi_{i}$ a collection of parameters that are common to all blocks. Further, $G_{i}(\,\,\cdot\mid\bxi_{i})$ is a suitable prior for $\btheta_{i}$. The data generating mechanism \eqref{Data.distribution} along with prior distributions for $\rho_{i}$ and $\bxi_{i}$ (if applicable) completely specify the Bayesian change point PPM.  

It is common to select $\Ls_{i}(\,\,\cdot\mid\btheta_{i},\bxi_{i})$ and $G_{i}(\,\,\cdot\mid\bxi_{i})$ such that they form a conjugate pair which results in $\Fs_{i}(\by_{i,j}\mid\bxi_{i})$ being available in closed form. That said, the choices for $\Ls_{i}(\,\,\cdot\mid\btheta_{i},\bxi_{i})$ and $G_{i}(\,\,\cdot\mid\bxi_{i})$ in \eqref{Data.distribution} can be quite general, depending on the nature of $\by_{i}$ and $\btheta_{i}$. Examples of this are data that follow an Ornstein-Uhlenbeck process with a Normal-Gamma prior for mean-precision parameters \citep{Martinez2014} (here, $\bm{\xi}_i$ is a case dependency parameter with a $\Uni(0,1)$ prior) and independent data belonging to the exponential family with a conjugate prior for the natural parameters \citep{Loschi2005} (in this case, there is no $\bxi_{i}$). The types of marginal likelihoods just described (and others not listed) are easily applied using our method. Even so, in what follows, we will focus on the following specification (which is suitable for changes in mean and variance for data supported on $\R$). Let $\btheta_{i}=(\mu_{i},\sigma_{i}^{2})^\top\in\bTheta_{i}=\R\times(0,+\infty)$ and
\begin{align}
    \begin{split}
        \Ls_{i}(\by_{i,j}\mid\btheta_{i})&=
        \prod_{t\in S_{i,j}}\Nor(y_{i,t}\mid\mu_{i},\sigma_{i}^{2}),
        \\
        G_{i}(\btheta_{i})&=
        \Nor(\mu_{i}\mid\mu_{i,0},(\kappa_{i,0})^{-1}\sigma_{i}^{2})
        \IGa(\sigma_{i}^{2}\mid\alpha_{i,0},\beta_{i,0}).
    \end{split}
    \label{Normal.NIG}
\end{align}
Here, $\mu_{i,0}\in\R$ and $\kappa_{i,0},\alpha_{i,0},\beta_{i,0}>0$ are fixed
hyperparameters, $\Nor(\,\,\cdot\mid\mu,\sigma^{2})$ denotes a normal density with
mean $\mu\in\R$ and variance $\sigma^{2}>0$, and $\IGa(\,\,\cdot\mid\alpha,\beta)$ denotes an inverse Gamma density with shape $\alpha>0$ and scale $\beta>0$. It is well known that the $j$th {\it data factor} induced by \eqref{Normal.NIG} is
given by
\begin{equation}
    \Fs_{i}(\by_{i,j})=t_{n_{i,j}}(\by_{i,j}\mid
    2\alpha_{i,0},\mu_{i,0}\bOne_{n_{i,j}},(\alpha_{i,0})^{-1}
    \beta_{i,0}\{\mI_{n_{i,j}}+(\kappa_{i,0})^{-1}\mJ_{n_{i,j}}\}),
    \label{Data.factor}
\end{equation}
where $n_{i,j}$ is the cardinality of $S_{i,j}$, $\bm{1}_{p}\in\R^{p}$ is the vector with entries equal 1, $\mI_{p}\in\R^{p\times p}$ is the identity matrix and $\mJ_{p}\in\R^{p\times p}$ is the matrix with all entries equal 1. Also, $t_{p}(\,\,\cdot\mid\nu,\bmu,\bSigma)$ is the $p$-dimensional Student's $t$-density with degrees of freedom $\nu>0$, location vector $\bmu\in\R^{p}$ and scale matrix $\bSigma\in\Sy^{p\times p}$, where $\Sy^{p\times p}$ denotes the space of positive-definite matrices.

The set of hyperparameters $(\mu_{i,0}$, $\kappa_{i,0}$,$\alpha_{i,0}$, $\beta_{i,0})^\top$ play a crucial role in determining what constitutes a change point. For example, setting $\alpha_{i,0}$ close to one will result in \eqref{Data.factor} having thick tails so that a change point would necessarily need to be far from the center. Conversely, with a large value of $\alpha_{i,0}$ \eqref{Data.factor} approximates a normal distribution and points not far from the center can still be change points. Consequently, thought must be dedicated to assigning values to the marginal likelihood parameters. In Section \ref{sec:tunning.parms} we discuss an empirical Bayes method that produces reasonable values for them in absence of prior information.


\section{The Joint Prior Distribution on a Collection of Partitions}\label{JointPriorDistribution}

We now describe our approach of formulating a joint model for a sequence of partitions. As mentioned, the partition $\rho_{i}$ for the $i$-th time series $\by_{i}$ has a one-to-one correspondence with $\bc_{i}$. Thus, any prior distribution for $\mC=(\bc_{1},\ldots,\bc_{L})$, say $\pi(\mC)$, uniquely determines a prior for $\brho=(\rho_{1},\ldots,\rho_{L})$. We start by describing our joint model as an extension of the change point PPM and then we connect it to $\pi(\mC)$, which is what we ultimately use in our approach as it facilitates computation. 

In our setup, rather than consider a single probability parameter $p_{i}$ for the $i$-th series, we define $\bm{\tilde{p}}_{i}=(p_{i,1},\ldots,p_{i,n-1})^\top$ with $p_{i,t}\in[0,1]$ and extend the cohesion in \eqref{cohesion.Yao} to
\begin{equation}
    c^{\star}(S_{i,j}\,;\,\bm{\tilde{p}}_{i})=
    \left\{
    \begin{array}{lc}
        p_{i,\tau_{i,j}}\prod_{t=\tau_{i,j-1}+1}^{\tau_{i,j}-1}(1-p_{i,t}), & \mathrm{if}\quad\tau_{i,j}<n,  \\
         \prod_{t=\tau_{i,j-1}+1}^{\tau_{i,j}-1}(1-p_{i,t}), & \mathrm{if}\quad\tau_{i,j}=n.
    \end{array}
  \right.
  \label{cohesion.Ours}
\end{equation}
Using the cohesion \eqref{cohesion.Ours} for contiguous partitions still results in $\sum_{\rho_{i}\in\Cs_{n}}\prod_{j=1}^{k_{i}}c^{\star}(S_{i,j}\,;\,\bm{\tilde{p}}_{i})=1.$
Therefore, the partition probabilities become
\begin{equation}
    \Pb(\rho_{i}=\{S_{i,1},\ldots,S_{i,k_{i}}\}\mid\bm{\tilde{p}}_{i})=
    \prod_{j=1}^{k_{i}}c^{\star}(S_{i,j}\,;\,\bm{\tilde{p}}_{i})=
    \prod_{t\in\bm{T}_{i}}p_{i,t}\prod_{t\notin\bm{T}_{i}}(1-p_{i,t}),
    \nonumber
\end{equation}
where $\bm{T}_{i}=\{\tau_{i,1},\ldots,\tau_{i,k_{i}-1}\}$. Including all $L$ partitions, the joint partition model
becomes
\begin{eqnarray*}
    \Pb(\rho_{i}=\{S_{i,1},\ldots,S_{i,k_{i}}\}:i=1,\ldots,L\mid\bm{\tilde{p}}_{1},\ldots,\bm{\tilde{p}}_{L})&=
    \prod_{i=1}^{L}\Pb(\rho_{i}=\{S_{i,1},\ldots,S_{i,k_{i}}\}\mid\bm{\tilde{p}}_{i}) 
    \nonumber \\
    &=\prod_{i=1}^{L}\left\{\prod_{t\in\bm{T}_{i}}p_{i,t}
    \prod_{t\notin\bm{T}_{i}}(1-p_{i,t})\right\}
    \nonumber
\end{eqnarray*}


It is important to stress that our extension ({\it i.e.}, introducing $p_{i,t}$) provides more flexibility for modeling simultaneous change point configurations $\brho$ by allowing us to correlate probabilities of a change point at each time point. As a consequence, the change point indicators $c_{i,t}$ are assumed conditionally independent with their own probability $p_{i,t}\in(0,1)$ of detecting a change $(c_{i,t}=1)$. 

Let $\bp_{t}=(p_{1,t},\ldots,p_{L,t})^\top$ be a $L$-dimensional vector of probabilities supported on the space $(0,1)^{L}$.  
To specify a multivariate distribution for $\bp_{t}$, we consider the bijective transformation  $\logit(\bp_{t})=\big(\log\big(\frac{p_{1,t}}{1-p_{1,t}}\big),\ldots, \log\big(\frac{p_{L,t}}{1-p_{L,t}}\big)\big)^\top$ which is defined on the Euclidean space $\R^{L}$, and model it with a multivariate Student's-$t$ distribution. The reason for selecting a Student's-$t$ distribution instead of, for example, a multivariate normal distribution is that extreme probabilities (near 0 or 1) are more achievable due to the thicker tails of the Student's-$t$. In summary, the proposed model for partitions $\brho$ can be formulated using the following hierarchical structure:
\begin{eqnarray} 
\begin{split}
c_{i,t}\mid p_{i,t}&\stackrel{\ind}{\sim}p_{i,t}^{c_{i,t}}(1-p_{i,t})^{1-c_{i,t}},\\
\logit(\bp_{t})&\stackrel{\iid}{\sim} t_{L}(\nu_{0},\bmu_{0},\bSigma_{0}),
\label{Prior.CP.indicators}
\end{split}
\end{eqnarray}
with $\bmu_{0}=(\mu^{0}_{1},\ldots,\mu^{0}_{L})^\top$, $\bSigma_{0}=(\sigma^{0}_{l_{1},l_{2}}:l_{1},l_{2}\in\{1,\ldots,L\}).$ In what follows, we will refer to the model comprised of \eqref{Normal.NIG}, \eqref{cohesion.Ours}, and \eqref{Prior.CP.indicators} as the correlated change point product partition model or simply, CCP-PPM.

Note that, specifying adequate values for $\nu_{0}$, $\bmu_{0}$ and $\bSigma_{0}$ in \eqref{Prior.CP.indicators} must be done with caution as $p_{i,t}$ are not invariant to their selection. In the absence of information regarding these parameters, we provide an interesting approach for selecting them in Section \ref{sec:tunning.parms}. 

\subsection{Properties of the Joint Model on Partitions}\label{properties}

In this section, we provide some interesting properties that are consequences of modeling the
change point indicators with \eqref{Prior.CP.indicators}. The proofs of all propositions are provided in the Appendix. 

\begin{proposition} \label{prop.1}
    Under the assumptions in \eqref{Prior.CP.indicators}, each of the $\bc_{i},$ $i=1,\ldots,L,$
    follows a change point PPM based on Yao's cohesion with probability parameter
    \begin{equation}
        \phi_{i}=\int_{\R}\left\{\frac{\exp(z)}{1+\exp(z)}\right\}
        t_{1}(z\mid\nu_{0},\mu^{0}_{i},\sigma^{0}_{i,i})\diff z.
        \label{Phi}
    \end{equation}
\end{proposition}

\noindent A consequence of Proposition \ref{prop.1} is that the correlation in the change point probabilities from our model only exists across the $L$ series for a fixed $t$. Within a series, the probability of a change point at time $t_{1}$ is independent of time $t_{2}$. 

The next proposition provides an interesting result regarding the number of expected change points based on the CCP-PPM.

\begin{proposition} \label{prop.2}
    Under the assumptions in \eqref{Prior.CP.indicators}, the number of
    change points $(k_{i}-1)$ for the $i$th series satisfies
    $(k_{i}-1)\sim\Bin(n-1,\phi_{i})$, where $\phi_{i}$ is given by
    \eqref{Phi}. Additionally, for series $i$ and $s$, the distribution for
    $(k_{i}-1,k_{s}-1)^\top$ is a mixture of a product of two Poisson's Binomial
    distributions \citep{Wang1993} with
    $\Cov(k_{i}-1,k_{s}-1)=(n-1)(\varphi_{i,s}-\phi_{i}\phi_{s})$, where
    \begin{equation}
        \varphi_{i,s}=\int_{\R^{2}}
        \left[\frac{\exp(z_{i})\exp(z_{s})}{\{1+\exp(z_{i})\}\{1+\exp(z_{s})\}}\right]
        t_{2}(\bz_{A}\mid\nu_{0},\bmu_{A,0},\bSigma_{A,0})
        \diff\bz_{A}.
        \label{Varphi}
    \end{equation}
    Here, $\bm{z}_{A}=(z_{i},z_{s})^{\top}$, $\bm{\mu}_{A,0}=(\mu^{0}_{i},\mu^{0}_{s})^{\top}$ and
    $\bSigma_{A,0}=(\sigma^{0}_{l_{1},l_{2}}:l_{1},l_{2}\in \{i,s\})$.
\end{proposition}

\begin{figure}[htbp]
  \begin{center}
    \includegraphics[scale=0.6]{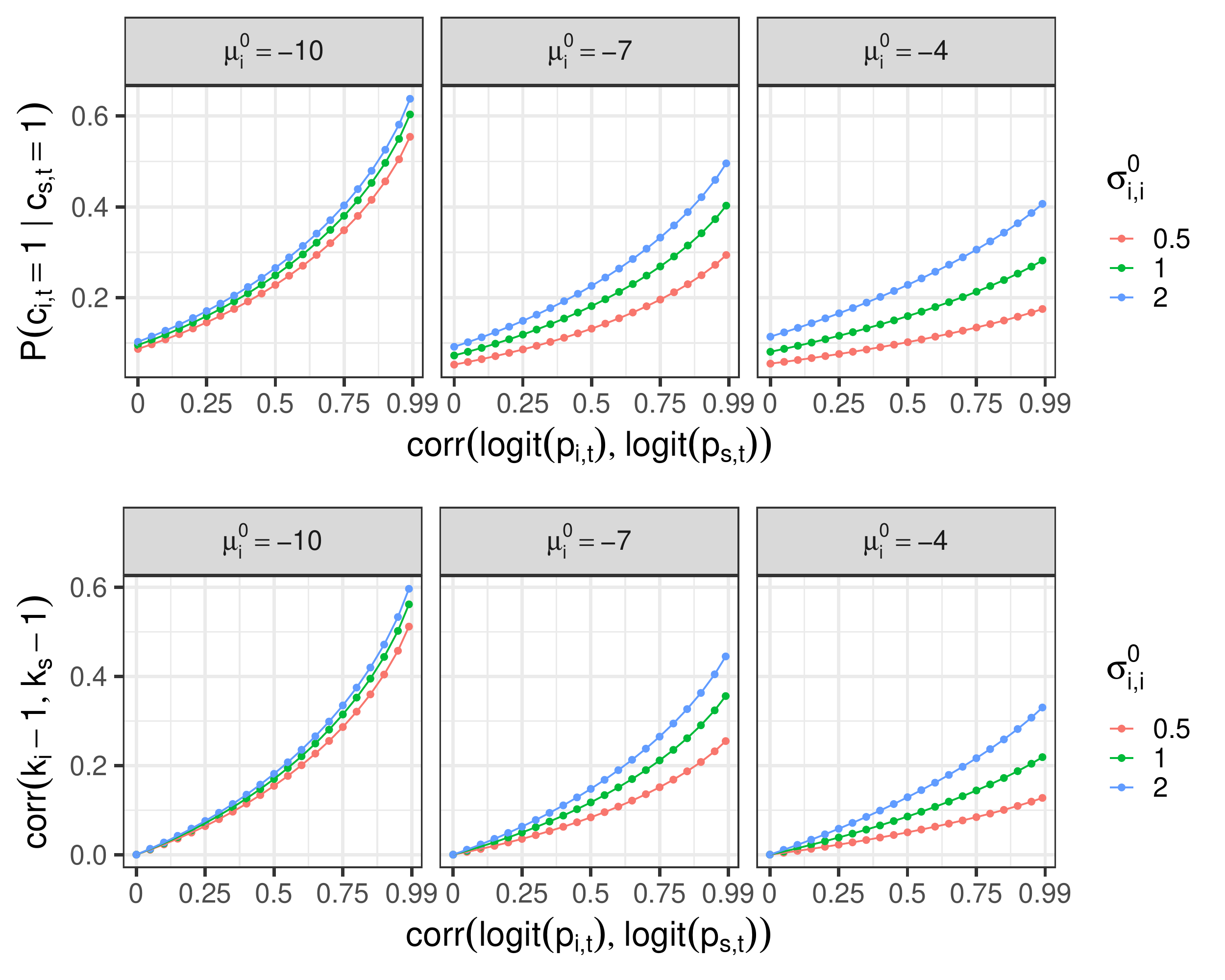}
  \end{center}    
    \caption{The top row displays values of \ref{cond.prob} for a few values of $\mu_i^0$ and $\sigma_i^0$ and in increasing sequence of correlations between  $\logit(p_{i,t})$ and $\logit(p_{s,t})$. The bottom row displays $\Corr (k_{i}-1, k_{s}-1)$ for the same values of $\mu_i^0$, $\sigma_i^0$, and correlations between  $\logit(p_{i,t})$ and $\logit(p_{s,t})$. The integral in \eqref{Varphi} was approximated numerically.}
    \label{fig:cond.prob}
\end{figure}

\noindent Note that $\E(k_{i}-1) = (n-1)\phi_{i}$ easily follows from   Proposition \ref{prop.2}.  For the values of $\mu^{0}_{i}$ and $\sigma^{0}_{i,i}$ considered in Figure \ref{fig:cond.prob}, the expected number of change points in each series ranges between 6.19 and 12.29 {\it a priori}. The bottom row of Figure \ref{fig:cond.prob} provides values for $\Corr (k_{i}-1, k_{s}-1)$ based on a numerical approximation of $\varphi_{i,s}$. As expected the number of change points in two series with high correlation between $\logit(p_{i,t})$ and $\logit(p_{s,t})$ will be similar. 

The last proposition derives the conditional probabilities of change point indicators. These are of particular interest as they illustrate how the probability of a change point across series varies as one series experiences a change point.

\begin{proposition}
    Under the assumptions in \eqref{Prior.CP.indicators}, the probability of a
    change point occurring at time $t$ in the $i$th series given that one occurred
    at time $t$ in the $s$th series is
    \begin{equation}\label{cond.prob}
        \Pb(c_{i,t}=1\mid c_{s,t}=1)=\frac{\varphi_{i,s}}{\phi_{s}}.
    \end{equation}
    Here $\phi_{s}$ and $\varphi_{i,s}$ are given by \eqref{Phi} and \eqref{Varphi},
    respectively.
\end{proposition}
\noindent The top row of Figure \ref{fig:cond.prob} provides values for $\Pb(c_{i,t}=1\mid c_{s,t}=1)$. The integral in $\Pb(c_{i,t}=1\mid c_{s,t}=1)$ was approximated using the statistical software {\tt R} (\citealt{R}). As expected, the higher the correlation between $\logit(p_{i,t})$ and $\logit(p_{s,t})$ the higher the conditional probabilities {\it a priori}, for the values of $\mu^{0}_{i}$ and $\sigma^{0}_{i,i}$ considered.

\subsection{Selection of tuning parameters} \label{sec:tunning.parms}

Like most change point methods, the posterior probability of classifying a point as a change point can be sensitive to marginal likelihood specification and prior parameter selection. We refer to these parameters as tuning parameters. In some cases, the practitioner can inform the procedure regarding a change point, which guides tuning parameter selection. Without this information, it is appealing to have a procedure that produces {\it default} values for the tuning parameters. 
Therefore, we describe an empirical Bayes approach to selecting values for
$(\mu_{i,0}$, $\kappa_{i,0}$, $\alpha_{i,0}$, $\beta_{i,0})^\top$. The approach we describe is geared towards situations in which the magnitudes of change points are relatively small.
%

First note that from \eqref{Data.factor} we have
\begin{equation}
    0\leq\Corr(y_{i,r},y_{i,s})=\frac{\Pb(e_{i,r}=e_{i,s})}{1+\kappa_{i,0}}\leq\frac{1}{1+\kappa_{i,0}},
    \nonumber
\end{equation}
for all $r,s\in\{1,\ldots,n\}$ and $r\neq s$. Under the scenario of no change points, $\Corr(y_{i,r},y_{i,s})$ is equal to the upper bound $(1+\kappa_{i,0})^{-1}$, which is constant as a function of $(r,s)$. Thus, a value for $\kappa_{i,0}$ could be empirically selected using $\Corr(y_{i,r},y_{i,s})$.  This correlation can be estimated by inspecting the empirical $\ell$-lag autocorrelation for $\by_{i}$, say $\widehat{c_{i,\ell}}$, and choosing the smallest $\ell\geq1$ such that $\widehat{c_{i,\ell}}>0$.  Then set $\kappa_{i,0}=(\widehat{c_{i,\ell}})^{-1}(1-\widehat{c_{i,\ell}})$. From there, the maximum likelihood estimators of the parameters in the Student's $t$-distribution given in \eqref{Data.factor} based on the entire $\by_{i}$ can be used to provide values for $\mu_{i,0}$, $\alpha_{i,0}$, and $\beta_{i,0}$. Let $d=2\alpha_{i,0}$, $m=\mu_{i,0}$ and $s=(\alpha_{i,0}\kappa_{i,0})^{-1/2}\{\beta_{i,0}(\kappa_{i,0}+1)\}^{1/2}$. Then, we set $\mu_{i,0}=\widehat{m_{i}}$, $\alpha_{i,0}=0.5\widehat{d_{i}}$ and $\beta_{i,0}=0.5\widehat{d_{i}}(1-\widehat{c_{i,\ell}})\widehat{s_{i}}^{2}$ where $(\widehat{d_{i}},\widehat{m_{i}},\widehat{s_{i}})
^\top$ denote the maximum likelihood estimates of $(d,m,s)^\top$.

Now, we focus on $(\nu_{0},\bmu_{0},\bSigma_{0})^\top$. Although $\E(\bp_{t})$ and $\Var(\bp_{t})$ do not exist in a closed form, a first-order Taylor expansion provides approximations to them. If $\nu_{0}>2$, then
\begin{eqnarray*}
    \E(\bp_{t})&\approx&\left(\frac{\exp(\mu^{0}_{1})}{1+\exp(\mu^{0}_{1})},\ldots,
    \frac{\exp(\mu^{0}_{L})}{1+\exp(\mu^{0}_{L})}\right)^\top,
    \nonumber \\
    \Var(\bp_{t})&\approx&\left(\frac{\nu_{0}}{\nu_{0}-2}\right)\mJ(\bmu_{0})\bSigma_{0}
    \mJ(\bmu_{0}),
    \nonumber \\
    \mJ(\bmu_{0})&=&\diag\left(\frac{\exp(\mu^{0}_{1})}{\{1+\exp(\mu^{0}_{1})\}^{2}},
    \ldots,\frac{\exp(\mu^{0}_{L})}{\{1+\exp(\mu^{0}_{L})\}^{2}}\right).
    \nonumber
\end{eqnarray*}
After choosing prior guesses for $\E(\bp_{t})$ and $\Var(\bp_{t})$, say $\bm{m}_{0}=(m^{0}_{1},\ldots,m^{0}_{L})^\top$ and $\mS_{0}$ respectively, we set
\begin{eqnarray*}
    \bmu_{0}&=&\left(\log\left(\frac{m^{0}_{1}}{1-m^{0}_{1}}\right),\ldots,
    \log\left(\frac{m^{0}_{L}}{1-m^{0}_{L}}\right)\right)^\top,
    \nonumber \\
    \bSigma_{0}&=&\left(\frac{\nu_0-2}{\nu_0}\right)\mD(\bm{m}_{0})^{-1}\mS_{0}\mD
    (\bm{m}_{0})^{-1},
    \nonumber \\
    \mD(\bm{m}_{0})^{-1}&=&\diag\left(\frac{1}{m^{0}_{1}(1-m^{0}_{1})},\ldots,
    \frac{1}{m^{0}_{L}(1-m^{0}_{L})}\right).
    \nonumber
\end{eqnarray*}
We recommend setting $\nu_{0}=3$, which is the least integer such that the approximations described above exist. In the case that no prior information is available to guide specifying $\bm{m}_{0}$ and $\mS_{0}$, the following empirical approach can be used.  Set $m^{0}_{1}=\cdots=m^{0}_{L}=n^{-1}$. For $\mS_{0}$, a compound symmetry covariance matrix $\sigma^{2}_{0}\{(1-r_{0})\mI_{L}+r_{0}\mJ_{L}\}$ can be used, where $\sigma^{2}_{0}= m^{0}_{1}(1-m^{0}_{1})/n = n^{-3}(n-1)$ and $r_{0}=0.5$.

\subsection{Posterior Sampling}\label{sampling}

The joint posterior distribution of $\bm{p}_t$ and $\mC$ is not analytically tractable. Therefore we resort to sampling from it using Markov Chain Monte Carlo (MCMC) methods. The MCMC algorithm we construct is very straightforward to implement and is an extension of the Gibbs sampler described in \cite{Loschi2003} with the main difference being that we employ Metropolis updates when conditional conjugacy is not available. Before model fitting, we recommend scaling each series to have mean zero and standard deviation one. The full conditional distributions we need in our algorithm are described next.

\begin{itemize}
    \item For each $i\in\{1,\ldots,L\}$ and $t\in\{1,\ldots,n-1\}$ update 
    $p_{i,t}$ according to its full conditional density $\pi(p_{i,t}\mid\cdots)$,
    which is proportional to
    \begin{equation}
        \frac{1}{p_{i,t}}
        \left(\frac{p_{i,t}}{1-p_{i,t}}\right)^{c_{i,t}}
        \left\{1+\frac{(\logit(\bp_{t})-\bmu_{0})'\bSigma_{0}^{-1}
        (\logit(\bp_{t})-\bmu_{0})}{\nu_{0}}\right\}^{-\frac{\nu_{0}+L}{2}}
        1(p_{i,t}\in(0,1)).
        \nonumber
    \end{equation}
    Here, $1(\,\,\cdot\,\in S)$ is the indicator function of the set $S$. To update $p_{i,t}$, we employ a random 
    walk Metropolis step with a normal centered at the previous iteration's value as a candidate density.
    The standard deviation of the normal candidate density is set to 0.005 which produces an acceptance rate in the general range of 0.2 and 0.5
  
    \item For each $i\in\{1,\ldots,L\}$, $t\in\{1,\ldots,n-1\}$ and $a\in\{0,1\}$,
    define the set of change point indicators
    $\bc^{\sss{(a)}}=(c_{1}^{\sss{(a)}},\ldots,c_{n-1}^{\sss{(a)}})^
    {\top}$ such that
    \begin{equation}
        c_{s}^{\sss{(a)}}=
        \left\{
            \begin{array}{ll}
                c_{i,s}, & \mbox{if $s\neq t$} \\
                a, & \mbox{if $s=t$.}
            \end{array}
        \right.
        \nonumber
    \end{equation}
    Using $\bc^{\sss{(a)}}$, we construct the corresponding set of cluster labels
    $\be^{\sss{(a)}}=(e_{1}^{\sss{(a)}},\ldots,e_{n}^{\sss{(a)}})
    ^{\top}$. Then, after computing
    \begin{equation}
        \varpi_{i,t}=
        \frac{\Pb(c_{i,t}=1\mid\cdots)}{\Pb(c_{i,t}=0\mid\cdots)}=
        \frac{\Fs_{i}(e_{t}^{\sss{(1)}}\mid\bxi_{i})
        \Fs_{i}(e_{t+1}^{\sss{(1)}}\mid\bxi_{i})}
        {\Fs_{i}(e_{t}^{\sss{(0)}}\mid\bxi_{i})}
        \left(\frac{p_{i,t}}{1-p_{i,t}}\right),
        \nonumber
    \end{equation}
    where $\Fs_{i}(j\mid\bxi_{i})=\Fs_{i}(\by_{i,j}\mid\bxi_{i})$,   $c_{i,t}$ can be updated using a Bernoulli
    distribution with probability parameter
    \begin{equation}
        \Pb(c_{i,t}=1\mid\cdots)=\frac{\varpi_{i,t}}{1+\varpi_{i,t}}.
        \nonumber
    \end{equation}

\end{itemize}
Now, an MCMC algorithm can be obtained by cycling through each of the full conditionals individually. If a model is proposed so that $\bxi_{i}$ is available, it is relatively straightforward to update $\bxi_{i}$ in the Gibbs sampler using a Metropolis step. The update is based on the following full conditional of $\bxi_{i}$ for each $i\in\{1,\ldots,L\}$
\begin{equation}
    \pi(\bxi_{i}\mid\cdots)\propto
    \left\{\prod_{j=1}^{k_{i}}\Fs_{i}(\by_{i,j}\mid\bxi_{i})\right\}h_{i}(\bxi_{i}),
    \nonumber
\end{equation}
where $h_{i}$ is a prior density for $\bxi_{i}$.

\section{Simulation Study}\label{simulation}

We conduct a small numerical experiment to study the CCP-PPM's ability to detect multiple change points. The experiment is based on generating data sets containing change points whose times are dependent across series, mimicking the {\it contagion} idea. We consider change points that result from simultaneous changes in the mean and variance of a normal distribution. Data sets are generated using three scenarios, with each one producing data sets containing $L=2$ series of $n=100$ observations. The scenarios are detailed next.

\begin{figure}[htbp]
    \begin{center}
        \includegraphics[scale=0.9]{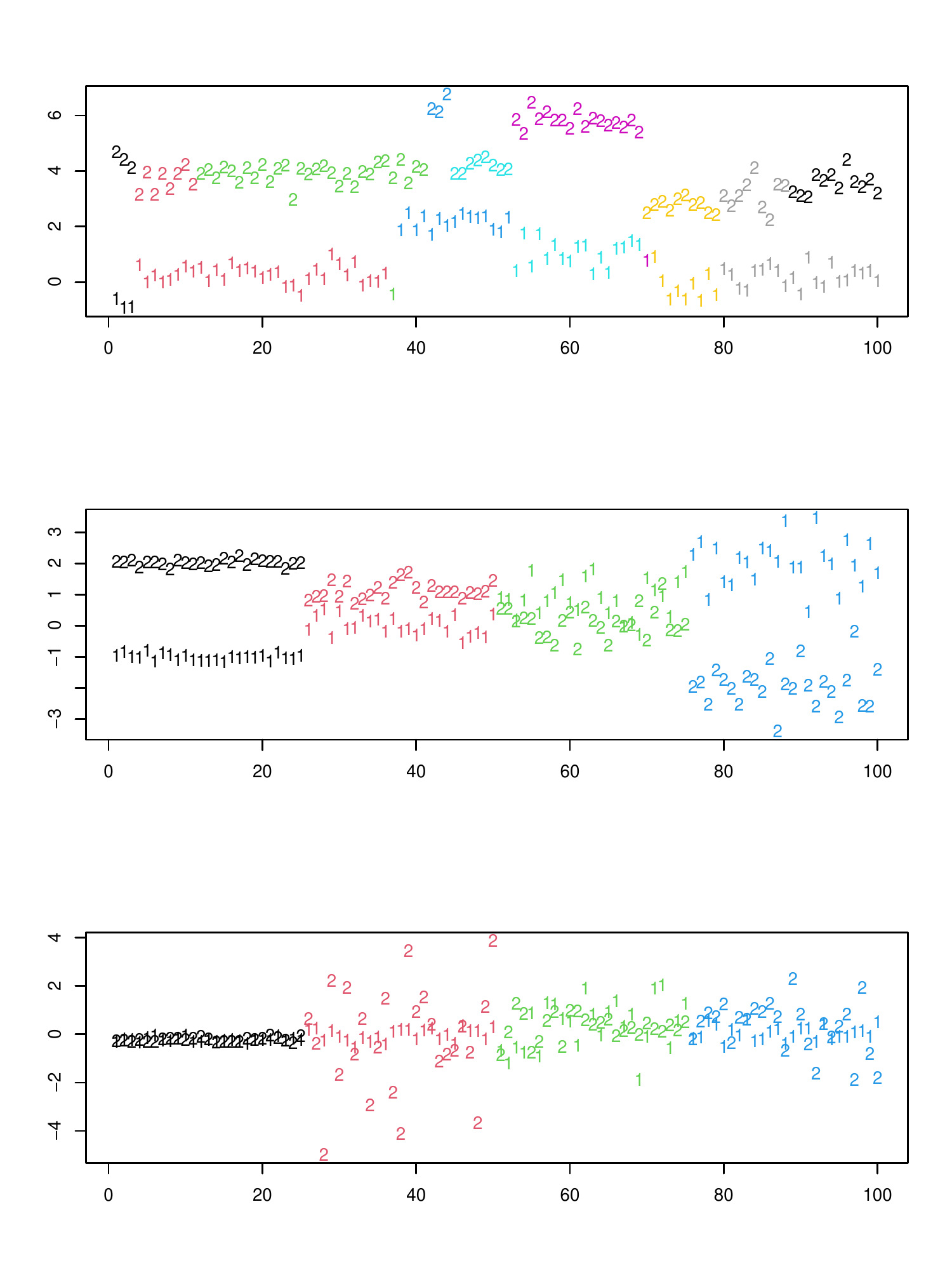}
        \caption{Example of the types of data sets used in the simulation study. Top row corresponds to {\it data type 1}, while the second and third rows to {\it data type 2} and {\it data type 3} respectively. Points that are labeled as {\bf 1} come from the first series with those labeled {\bf 2} come from the second. The different colors identify the clusters formed by change points.}
        \label{sim.study.data}
    \end{center}
\end{figure}

The first scenario referred as {\it data type 1} uses the CCP-PPM as a data generating mechanism. We set $\nu_0 = 3$, $\bm{\mu}_{0} = (-6, -6)^{\top}$, and $\bm{\Sigma}_0$ to a compound symmetric matrix with variance 10 and correlation 0.9. Based on these values, \eqref{Prior.CP.indicators} is used to create partitions. Once partitions are formed, we use a $\Nor(0,1)$ to generate cluster specific means for one series and a $\Nor(4,1)$ for the other.  Cluster specific variances were all generated using an $\IGa(10,1)$ distribution. Once cluster specific parameters were generated, observations were generated using a $\Nor(0,1)$. An example of this type of data is displayed in the top row of Figure \ref{sim.study.data}.

The following two scenarios set change point locations in both series at 25, 50, and 75. As a result, the change points of the two series are highly dependent. Under this setting, four clusters of 25 observations for both series are obtained. Given this type of partition, we produce observations in two ways. The first one, which we refer to as {\it data type 2}, generates observations using a normal distribution with the following cluster-specific means and variances:
\begin{enumerate}
\item[-] $\bmu^{\star} = (-1, 0, 1, 2)^{\top}$ and $\bsigma^{\star} = (0.1, 0.25, 0.5, 0.75)^{\top}$,
\item[-] $\bmu^{\star} = (2, 1, 0, -1)^{\top}$ and $\bsigma^{\star} = (0.1, 0.25, 0.5, 0.75)^{\top}$.
\end{enumerate}
The second scenario, which we refer to as {\it data type 3}, generates observations using a normal distribution and cluster-specific means and variances given by:
\begin{enumerate}
\item[-] $\bmu^{\star} = (-0.25,0,0.25,0.5)^{\top}$ and $\bsigma^{\star} = (0.1, 0.25, 1, 0.25)^{\top}$,
\item[-] $\bmu^{\star} = (-0.25,0,0.25,0.5)^{\top}$ and $\bsigma^{\star} = (0.1, 2, 0.5, 1)^{\top}$.
\end{enumerate}

This scenario is included because it provides insight to how the CCP-PPM approach performs for data similar to that which we consider in Section \ref{sec:application}. Examples of synthetic data sets created from these two scenarios are provided in the second and third rows of Figure \ref{sim.study.data}. It is important to note that in the scenario {\it data type 2}, the means are the primary driver of change points, while in the scenario {\it data type 3} the variances, or volatility, is the primary driver of change points.

We simulated one hundred data sets for each scenario. Then, we fit the CCP-PPM by collecting 2000 MCMC iterates after discarding the initial 10,000 as burn-in and thinning by 10 all of which was carried out using the {\tt ppmSuite} (\citealt{ppmSuite}) {\tt R} package that is available on {\tt CRAN}. In addition, we also consider the following controls for detecting change points:

\begin{itemize}
    \item The PPM-based method of \cite{wang2015bayesian}. Since this approach does not analyze multiple series, we fit it to each series independently instead. The {\tt bcp} package \citep{bcp_Rpackage} found in the {\tt R} statistical software \citep{R} is used to implement this method. We referred to this method as the {\it Wang} method.
    
    \item The method of \cite{matteson2014}. For this method, the two series are analyzed jointly. This method is implemented using the {\tt R} package {\tt ecp} \citep{ecp_Rpackage}. We referred to this method as the {\it Matteson} method.
    
    \item The method developed in \cite{Barry1992} and \cite{Loschi2002}. This method is our most natural competitor and is implemented using the {\tt ppmSuite} {\tt R} package.  This method is referred to as the {\it Loschi} method. As for the {\it Wang} method, the {\it Loschi} method is fit to each series independently.  
\end{itemize}

The tuning parameters used for the CCP-PPM are the following. For the marginal likelihood we use $(\mu_{i,0},\kappa_{i,0},\alpha_{i,0},\beta_{i,0})^{\top} = (0, 1, 2, 1)^{\top}$. These same values are used for the {\it Loschi} method.  For the prior on $\bm{p}_t$, we use the same values that were used to generate the {\it data type 1}.  For the {\it Loschi} method there is a single $p_i$ for each series and we used $p_i \stackrel{\iid}{\sim}\mbox{Beta}(1, 20)$. Default values of the {\tt bcp} and {\tt ecp} {\tt R}-packages were considered for the {\it Wang} and {\it Matteson} methods, respectively.  

\begin{figure}
    \centering
    \includegraphics[scale=0.75]{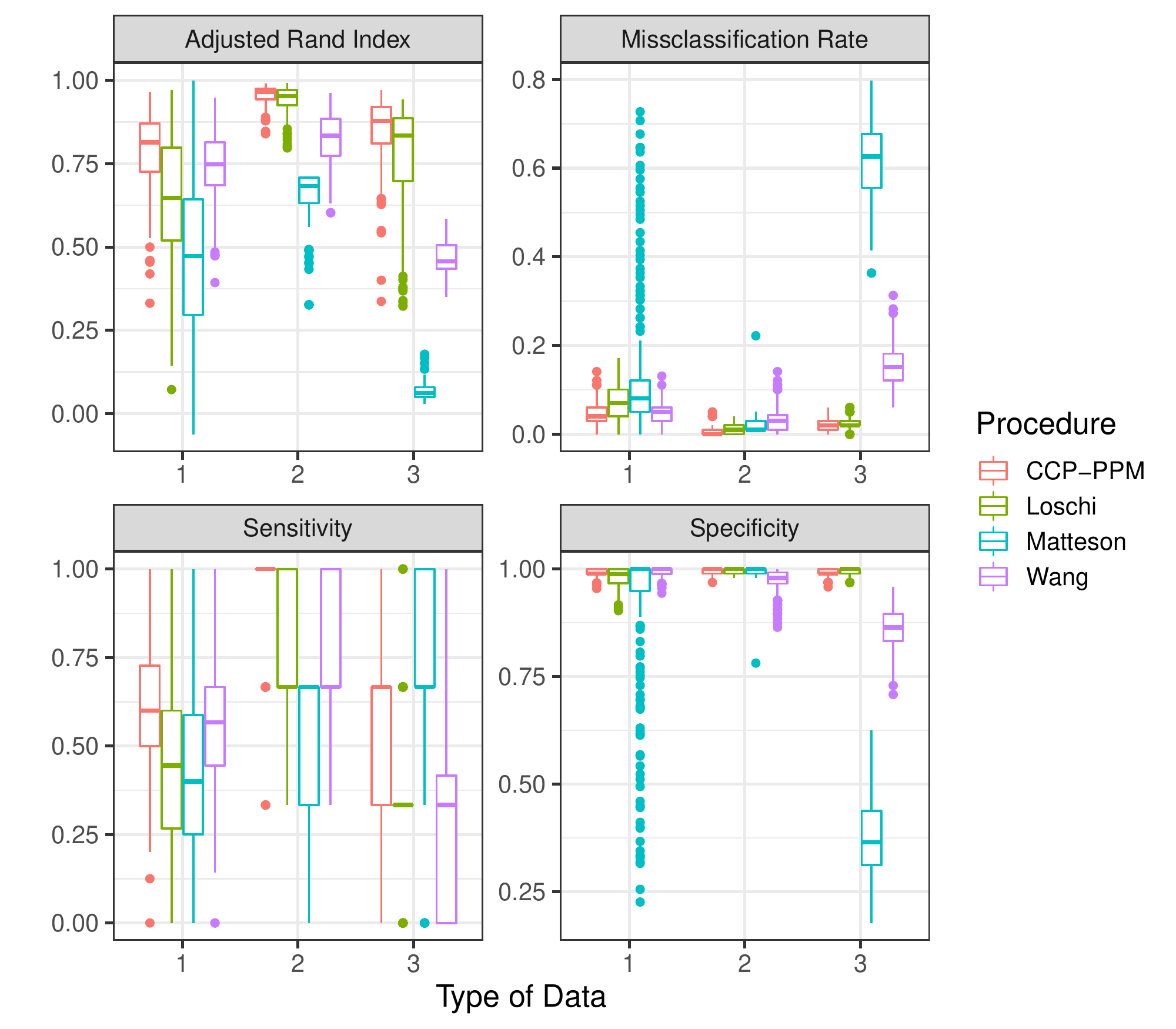}
    \caption{Results from the simulation study.  Each boxplot displays the results for each method based on the 100 datasets generated. For the adjusted Rand index, sensitivity (true positive rate), and specificity (true negative rate) higher values indicate superior performance.}
    \label{fig:sim_study}
\end{figure}

After considering the methods described above in each scenario, we classify any point as a change if its posterior probability of being a change is greater than 0.5. Then, we compute the overall misclassification rate, sensitivity (true positive rate), and specificity (true negative rate) of classifying points as change points. These metrics give us information on the accuracy of identifying points as change points. In addition, we calculate the adjusted Rand index \citep{ARI} (ARI) between the estimated partition based on change points and the true partition. This metric illustrates how well each method does at recovering the true partition of contiguous clusters.

The results of the simulation study are provided in Figure \ref{fig:sim_study}. Note that the CCP-PPM outperforms all the considered methods at recovering the true partition according to the adjusted Rand index regardless of the type of data. In terms of the overall misclassification rate, our method performs better than the {\it Loschi} and {\it Matteson} methods. The {\it Wang} method performs  comparably way to our approach in the scenario 1 ({\it data type 1}), but in scenario 3 ({\it data type 3}) this method performs poorly. Regarding the sensitivity, our approach outperforms the other three methods in the first and second scenarios ({\it data type 1} and {\it data type 2}, respectively). Nonetheless, in the third scenario ({\it data type 3}) {\it Matteson}  has the highest sensitivity but at the cost of very low specificity. In summary, it appears that the CCP-PPM overall performs the best at detecting the correct number and location of change points.

\section{Finance Data Application} \label{sec:application}

We now turn our attention to the application that motivated our proposal. As is commonly done in financial applications, we analyze returns rather than prices. Returns are defined as $R_{t}=(P_{t}-P_{t-1})/P_{t-1}$ for $t\in\{2,\ldots,n\}$, where $P_{t}$ is the daily price. When analyzing the data set, we consider {\it contagion} both between mature markets and emerging ones and also between emerging markets. We analyzed the most important Latin American markets (emerging markets), namely Argentinean, Brazilian, Chilean and Mexican markets, including the USA market in the analysis (mature market). Consequently, we fit the CCP-PPM and the change point PPM of Loschi or simply {\it Loschi} method (which is perhaps our method's most natural competitor, treating each series independently).  

We considered the return series of their main stock indexes, namely, the MERVAL (\'Indice de Mercado de Valores de Buenos Aires) of Argentina, the IBOVESPA (\'Indice da Bolsa de Valores do Estado de S\~ao Paulo) of Brazil, the IPSA (\'Indice de Precios Selectivos de Acciones) of Chile, the IPyC (\'Indice de Precios y Cotizaciones) of Mexico, and the Dow Jones (Dow Jones Industrial Average) of USA. The stock returns were recorded daily from October 31, 1995 to October 31, 2000.

We employ the procedure described in Section \ref{sec:tunning.parms} to produce values for the tuning parameters in \eqref{Data.factor} and \eqref{Prior.CP.indicators}. This resulted in values for $(\mu_{i,0},\kappa_{i,0},\alpha_{i,0},\beta_{i,0})^{\top}$ that are listed in Table \ref{marg.like.parms}. These tuning parameter values were used for both the CCP-PPM and that of {\it Loschi} method.

\begin{table}[ht!]
\begin{center}
\begin{tabular}{l cc cc} \toprule
Series      & $\mu_{i,0}$ & $\kappa_{i,0}$ & $\alpha_{i,0}$ & $\beta_{i,0}$ \\ \midrule
USA         &  0.009 & 188.924 & 2.212 & 1.233 \\
Mexico      & -0.014 &   9.075 & 1.596 & 0.574 \\
Argentina   &  0.009 &  10.186 & 1.328 & 0.402 \\
Chile       & -0.022 &   2.533 & 1.864 & 0.656 \\
Brazil      &  0.022 &  11.010 & 1.349 & 0.423 \\ \bottomrule
\end{tabular}
\end{center}
\caption{Values of $(\mu_{i,0},\kappa_{i,0},\alpha_{i,0},\beta_{i,0})^{\top}$ for each country's series. The values are the result of applying the empirical procedure described in \ref{sec:tunning.parms} to each country's returns.}
\label{marg.like.parms}
\end{table}

To specify values for $\bm{\mu}_0$ and $\bm{\Sigma_0}$, we first set $m^{0}_{1}=\cdots=m^{0}_{5}=n^{-1}$ (in the application $n=1309$) and used a compound symmetry matrix for $\mS_{0}$ with variance $\sigma^{2}_{0}= m^{0}_{1}(1-m^{0}_{1})/n = n^{-3}(n-1)$ and correlation $0.5$. This resulted in $\bm{\mu}_0 = (-7.1762)\bm{1}_{5}$ and $\bm{\Sigma}_0$ being a compound symmetric matrix with variance $0.334$ and correlation $0.5$. We set $\nu_0 = 3$.  For the {\it Loschi} method $p_i \sim \mbox{Beta}(a,b)$ and we set $a=1304.5$ and $b=681209.9$. These values were selected based on setting the mean number of clusters {\it a priori} to 3.5 with a variance of 2.5. 
Both methods were fit by collecting 1000 MCMC samples after discarding the first 10,000 as burn-in and thinning by 5 ({\it i.e.}, 15,000 total samples were collected). Both the CCP-PPM and {\it Loschi}'s method were fit  using the {\tt ccp\_ppm} and {\tt icp\_ppm} functions that are available in the {\tt ppmSuite}-package that can be found on {\tt CRAN}.  

There are two approaches that could used to estimate change points.  The first classifies points as change points if their posterior probability of being a change point is greater than some pre-specified value.  The second classifies points as change points based on a partition estimate.  We report both as both require input from the user (pre-specified probability cut-off for the first approach and a loss function in the second approach).

We first explore the {\it a posteriori} dependence between partitions from the five markets.  To do this, at each MCMC iteration we computed the ARI for all possible pairs of partitions (which is 10 in this application).  The CCP-PPM produced slightly more similar partitions across countries than the {\it Loschi} method.  The overall average pairwise ARI for CCP-PPM turned out to be 0.51 compared to 0.48 from the {\it Loschi} method.

\begin{figure}[htbp]
    \begin{center}
        \includegraphics[scale=0.66]{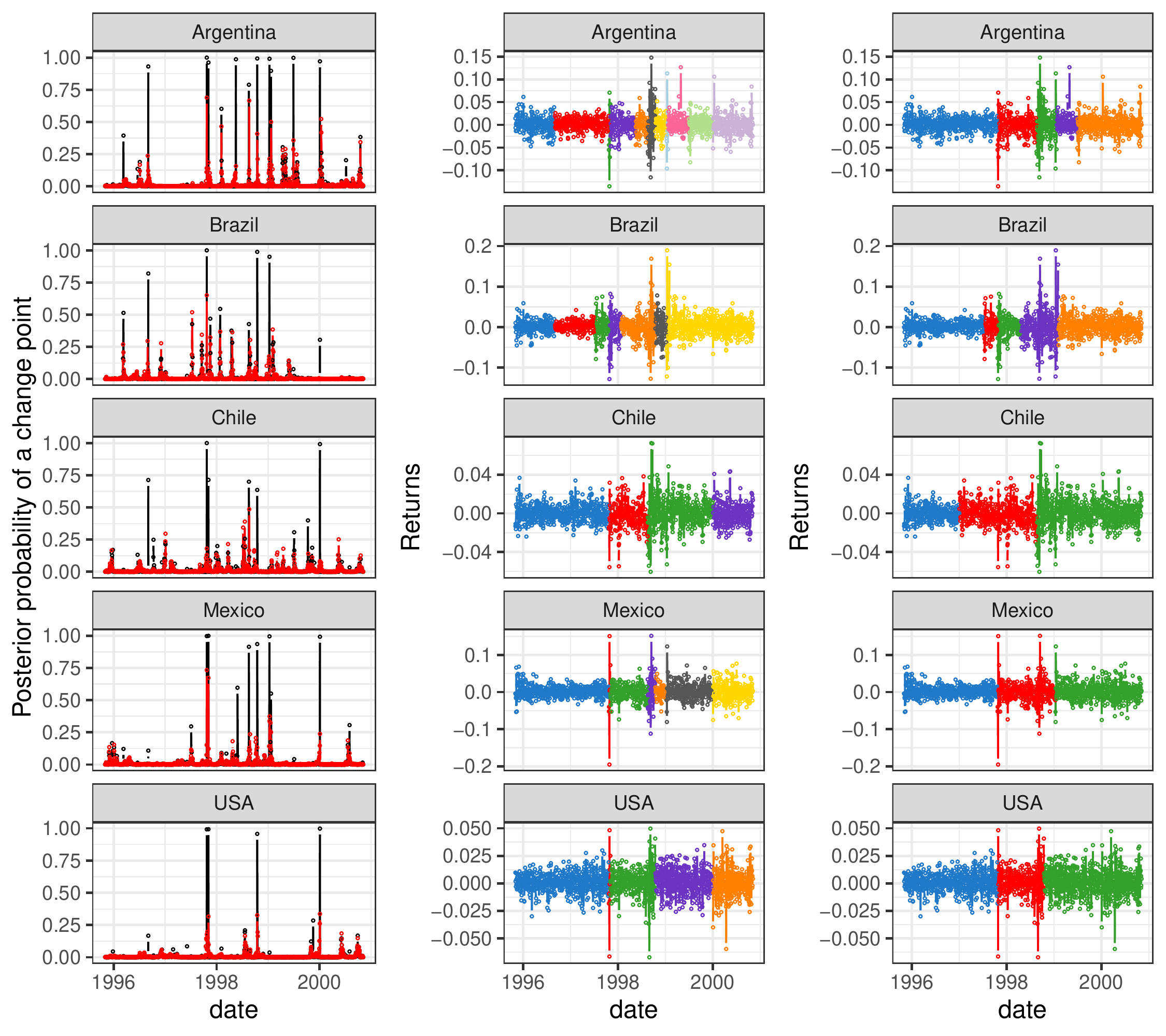}
        \caption{The left plot display posterior probability of each point being a change point with black points corresponding to the CCP-PPM method and red points to the {\it Loschi} method.  The middle plot displays the partition estimate of each series for the CCP-PPM and the right plot that for {\it Loschi}.  Both sets of partitions were estimated using the {\tt salso} package in {\tt R}.  }
        \label{app.fig.1}
    \end{center}
\end{figure}

Next we explore the posterior change point probabilities which are displayed in the first column of Figure \ref{app.fig.1}.  The black points correspond to the CCP-PPM and the red to {\it Loschi}.  For both, change point probabilities were estimated using the posterior means of $c_{i,t}$. It seems that there is a general agreement between the two methods regarding the location of potential change points. However, the CCP-PPM seems to produce probabilities that are closer to one for these points compared to the {\it Loschi} method.  In fact, the {\it Loschi} method never records a change point probability greater than 0.75.   Similarly, both methods agree on the general location of points that have a small chance of being a change point, although the CCP-PPM seems to push these probabilities closer to zero compared to the {\it Loschi} method.

Figure \ref{app.fig.1}, second and third columns, shows the partition estimates under the CCP-PPM (second column) and {\it Loschi} method (third column). Partition estimates were obtained using the {\tt salso} \citep{salso:2021} {\tt R} package and the generalization of the Variation of Information loss function \citep{VI2007}. Since in our case it seems natural to penalize change point false positives more than false negatives, we set the false positive penalty parameter of the {\tt salso} function to $a=25$ \citep[see][for more details]{dahl2021search}.  We note briefly that setting $a=25$ was driven primarily by {\it Loschi}'s method. If $a > 25$, then {\it Loschi}'s method tended to smooth over some change points and for $a< 25$ it tended to produce more change points than what would be desired. However, the CCP-PPM was reasonably robust to $a$'s value.  This is a consequence of the change point probabilities from the the {\it Loschi} method being more central (i.e., closer to 0.5) than those from the CCP-PPM. 

Apparent differences between the estimated partitions exist, and they illustrate how the CCP-PPM takes into account the dependency between the index series or the {\it contagion} phenomenon. For example, the CCP-PPM method identifies a shared partition for Brazil, Chile, Mexico, and the USA at the end of 1997. It is important to stress that in July 1997, the Thai government ran out of foreign currency, forcing it to float the Thai baht which is a factor in starting the 1997 Asian financial crisis, or Asian Flu. This crisis spread internationally, affecting some Asian stock markets such as Indonesia, South Korea, Hong Kong, Laos, Malaysia, Philippines, Brunei, mainland China, Singapore, Taiwan, and Vietnam. According to \cite{Harrigan2000}, the Asian crises' overall effect on the United States were small. However, as mentioned by \cite{Stalligns1998}, the Asian Flu hit the Latin American markets in October 1997, when bone spreads widened abruptly implying more risk. In the Argentinean market, our approach identified a first partition at the end of 1996. Note that, at the end of 1995, the (real) gross domestic product (GDP) in this country fell by 2.5 percent. However, by the end of 1996, it rebounded by 5.5 percent \citep{FMI2003}, possibly affecting the performance of the MERVAL index.

Another cluster or change point our method identifies is related to the Russian crisis or Russian Cold in August 1998. This crisis started when the Russian government and the Russian Central Bank devalued the ruble and defaulted on its debt. It is important to stress that although most countries experienced changes in their stock market returns series at the end of 1998, the Argentinean market experienced a change in mid-1998. Moreover, the IBOVESPA index (Brazil) experienced a change at the beginning of 1999, just after the Russian crisis. This crisis was known as the Samba effect and was produced when the Minas Gerais State Governor, Itamar Franco, stopped paying Minas Gerais debt to other states, generating unleashing capital flights. Note that a small cluster is detected by the CCP-PPM in the indexes of Argentina, Mexico, and USA towards the end of 1998, but something the {\it Loschi} method misses. These clusters provide some evidence that the {\it contagion} phenomena from a mature market to emerging ones is being captured by the CCP-PPM, which includes dependency between partitions.

Finally, the CCP-PPM also detected a cluster after the year 2000. The corresponding change point can be explained by the dot-com bubble, caused by excessive speculation of some internet companies in the late 1990s. On January 14, 2000, the Dow Jones Industrial Average reached its dot-com bubble peak. This partition is observed in the Argentinean, Chilean, Mexican, and USA markets. Possibly, the dot-com bubble may have affected other economies over the Latin American region, evidencing some {\it contagion} effect. In this case, {\it Loschi} method did not detect a change point at or near the above-mentioned date.



\begin{figure}[htbp]
    \begin{center}
        \includegraphics[scale=0.87]{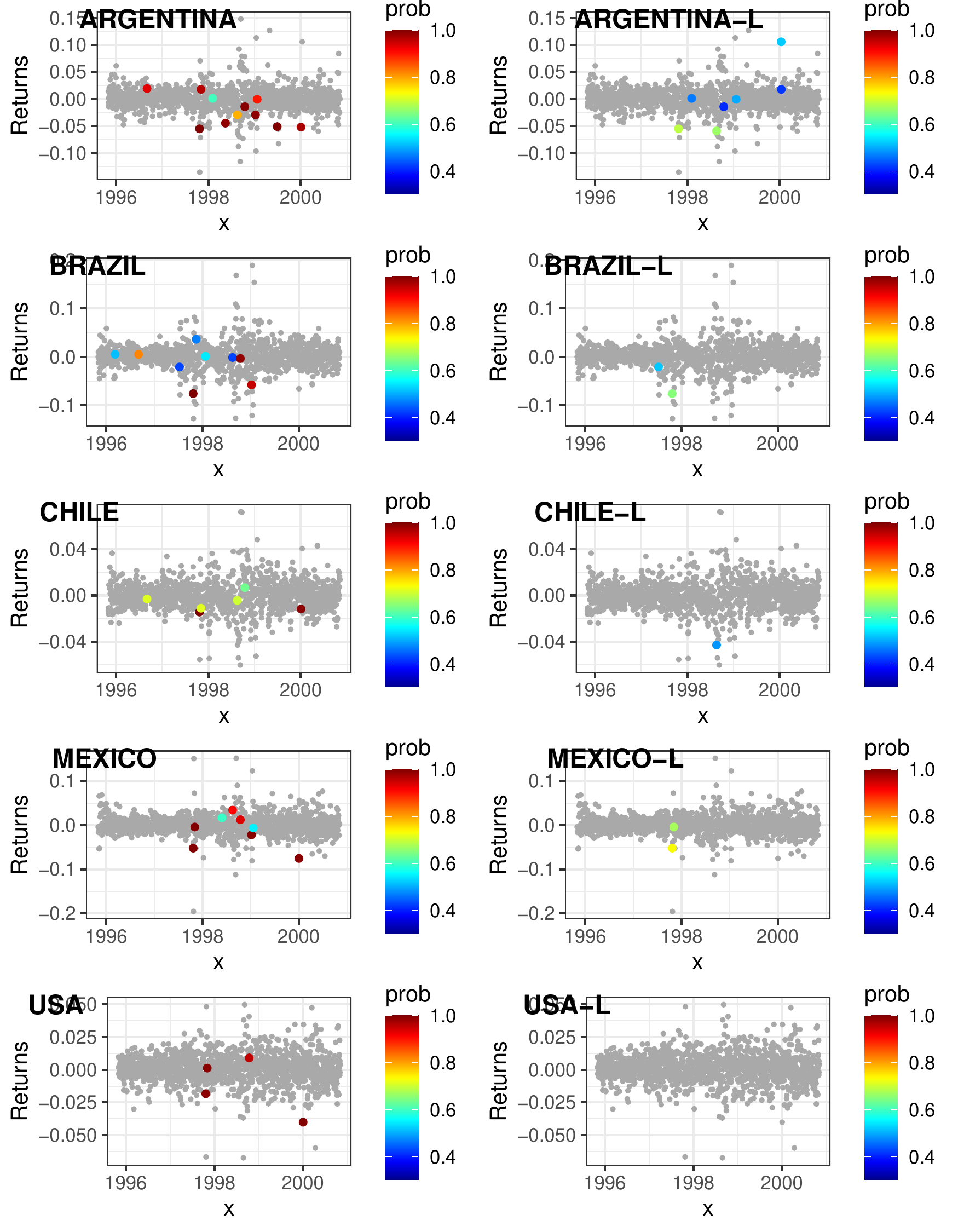}
        \caption{Change point posterior probabilities. Gray points correspond to locations whose posterior probability of being a change point was less than 0.4. The left column displays results under the CCP-PPM and the right column under the {\it Loschi} method. }
        \label{app.fig.4}
    \end{center}
\end{figure}

As mentioned, the estimated partitions in Figure \ref{app.fig.1} depend on the value $a=25$. To show a more complete picture of both methods performance, we provide Figure \ref{app.fig.4}. In this figure, all points with a posterior probability of being a change point less than 0.4 are colored gray. The left column corresponds to the CCP-PPM fit while the right the {\it Loschi} method. The CCP-PPM fit has more power in detecting change points compared to {\it Loschi} method, without inflating the false-positive rate. The points highlighted by the CCP-PPM fit are at least plausibly change points, and those associated with more pronounced volatility generally have a larger probability of being a change point, which is a desirable characteristic.

\section{Discussion}\label{discussion}

In this paper we developed a new change point detection model for $L$ time series with $n$ observations in an arbitrary space, which undergo sudden changes in their distributional parameters. By making dependent the vector of latent change point probabilities $\bm{p}_{t}=(p_{1,t},\ldots,p_{L,t})^{\top}$ at a specific time $t\in\{1,\ldots,n-1\}$, the corresponding $L$ partitions with contiguous clusters are encouraged to be correlated. We provide some theoretical results that help to better understand the main features of our model, a useful procedure to guide the specification of all parameters that are involved in, and simple pseudo-code to perform posterior inference via MCMC methods. Through a small simulation study, we compared the ability of our model with other compelling approaches to detect highly dependent changes under different scenarios, showing an improvement in detecting change points. Additionally, we applied our method to the returns of emerging Latin American and US markets, obtaining exciting results about possible {\it contagion effect} between the economies of these countries based on the dependence between change point locations.


In terms of extending the proposed model with the aim of making it more flexible, several directions can be pursued. For instance, the assumption that the vectors of change point probabilities $\bm{p}_{t}=(p_{1,t},\ldots,p_{L,t})^{\top}$ are independent and identically distributed through time $t$ can be relaxed. One possible approach would be to model $\bm{p}_{t}$ with a stationary process in $(0,1)^{L}$. Another interesting direction would be to incorporate time-dependent covariates in the marginal likelihood function to describe abrupt changes in a regression curve. A similar situation, but  more complex, is to incorporate covariates in the distribution for contiguous partitions. Finally, the computational cost involved in the MCMC algorithm for posterior inference increases rapidly as the length and number of time series grow. It would be appealing to develop strategies that mitigate the so-called ``curse of dimensionality''. These are all topics of future research.


\vspace{1cm}

\noindent {\bf Acknowledgements}

\noindent Jos\'e J. Quinlan gratefully recognizes the financial support provided by the Agencia Nacional de Investigaci\'on y Desarrollo de Chile (ANID) through Fondecyt Grant 3190324.

\section{Appendix}
\appendix

\section{Proofs of Propositions}
Proofs of the propositions depend on a well known result of the multivariate $t$.  For the sake of completeness, we include it here.

\begin{result}[\cite{Ding2016}]
    Let $A$ be a non-empty subset of $\{1,\ldots,L\}$.
    If $\bz=(z_{1},\ldots,z_{L})^{\top}\sim t_{L}(\nu_{0},\bmu_{0},\bSigma_{0})$, then
    $\bz_{A}=(z_{l}:l\in A)^{\top}\sim t_{|A|}(\nu_{0},\bmu_{A,0},\bSigma_{A,0})$, where
    $|A|$ is the cardinality of $A$, $\bmu_{A,0}=(\mu^{0}_{l}:l\in A)^{\top}$ and
    $\bSigma_{A,0}=(\sigma^{0}_{l_{1},l_{2}}:l_{1},l_{2}\in A)$. When $A=\{a\}$,
    set $z_{A}=z_{a}$, $\bmu_{A,0}=\mu^{0}_{a}$ and
    $\bSigma_{A,0}=\sigma^{0}_{a,a}$.
\end{result}

    \subsection*{Proof of Proposition 1.} A direct consequence of
    \eqref{Prior.CP.indicators} and Result 1 is that
    \begin{equation}
        \logit(p_{i,1}),\ldots,\logit(p_{i,n-1})\stackrel{\iid}{\sim}
        t_{1}(\nu_{0},\mu^{0}_{i},\sigma^{0}_{i,i}).
        \nonumber
    \end{equation}
    Invoking the Change of Variables theorem, the marginal distribution for 
    $\bc_{i}$ is given by
    \begin{align}
        \pi(\bc_{i})&=
        \E\left[\prod_{t=1}^{n-1}p_{i,t}^{c_{i,t}}(1-p_{i,t})^{1-c_{i,t}}\right]
        \nonumber \\
        &=\prod_{t=1}^{n-1}\E(p_{i,t})^{c_{i,t}}\{1-\E(p_{i,t})\}^{1-c_{i,t}}
        \nonumber \\
        &=\prod_{t=1}^{n-1}\phi_{i}^{c_{i,t}}(1-\phi_{i})^{1-c_{i,t}},
        \nonumber
    \end{align}
    where
    \begin{equation}
        \phi_{i}=\int_{\R}\left\{\frac{\exp(z)}{1+\exp(z)}\right\}
        t_{1}(z\mid\nu_{0},\mu^{0}_{i},\sigma^{0}_{i,i}).
        \nonumber
    \end{equation}
    Keeping in mind that $k_{i}=1+\sum_{t=1}^{n-1}c_{i,t}$,
    \begin{align}
        \pi(\bc_{i})&=
        \prod_{t=1}^{n-1}\phi_{i}^{c_{i,t}}(1-\phi_{i})^{1-c_{i,t}}
        \nonumber \\
        &=\phi_{i}^{k_{i}-1}(1-\phi_{i})^{n-k_{i}}.
        \nonumber
    \end{align}
    From the last expression, the result follows.
    
    \subsection*{Proof of Proposition 2.} From the proof of Proposition 1, we
    saw that $c_{i,1},\ldots,c_{i,n-1}$ are independent $\Ber(\phi_{i})$
    random variables. From this observation, it follows that
    \begin{equation}
        k_{i}-1=\sum_{t=1}^{n-1}c_{i,t}\sim\Bin(n-1,\phi_{i}).
        \nonumber
    \end{equation}
    As for the distribution of $(k_{i}-1,k_{s}-1)^{\top}$, given
    $\bm{p}_{i,s}=(p_{i,1},p_{s,1},\ldots,p_{i,n-1},p_{s,n-1})^{\top}$ the random 
    variables $(k_{i}-1)$ and $(k_{s}-1)$ are conditionally independent
    Poisson's Binomial with corresponding parameters 
    $\bm{\tilde{p}}_{i}=(p_{i,1},\ldots,p_{i,n-1})^{\top}$
    and $\bm{\tilde{p}}_{s}=(p_{s,1},\ldots,p_{s,n-1})^{\top}$. The mixture distribution
    follows since
    \begin{align}
        \Pb(k_{i}-1=a,k_{s}-1=b)&=
        \E\{\Pb(k_{i}-1=a,k_{s}-1=b\mid\bm{p}_{i,s})\}
        \nonumber \\
        &=\E\{\Pb(k_{i}-1=a\mid\bm{\tilde{p}}_{i})\Pb(k_{s}-1=b\mid\bm{\tilde{p}}_{s})\},
        \nonumber
    \end{align}
    for all $a,b\in\{0,\ldots,n-1\}$. Now, by the Law of Total Covariance
    \begin{align}
        \Cov(k_{i}-1,k_{s}-1)&=\Cov\left(\sum_{t_{1}=1}^{n-1}c_{i,t_{1}},
        \sum_{t_{2}=1}^{n-1}c_{s,t_{2}}\right)
        \nonumber \\
        &=\sum_{t_{1}=1}^{n-1}\sum_{t_{2}=1}^{n-1}\Cov(c_{i,t_{1}},c_{s,t_{2}})
        \nonumber \\
        &=\sum_{t_{1}=1}^{n-1}\sum_{t_{2}=1}^{n-1}\Cov(p_{i,t_{1}},p_{s,t_{2}}).
        \nonumber
    \end{align}
    Due to \eqref{Prior.CP.indicators} and Result 1 with its respective notation,
    \begin{equation}
        (\logit(p_{i,1}),\logit(p_{s,1}))^{\top},\ldots,
        (\logit(p_{i,n-1}),\logit(p_{s,n-1}))^{\top}
        \stackrel{\iid}{\sim}
        t_{2}(\nu_{0},\bm{\mu}_{A,0},\bm{\Sigma}_{A,0}),
        \nonumber
    \end{equation}
    where $A=\{i,s\}$. By the Change of Variables theorem and the previous observation, we
    have that
    \begin{equation}
        \Cov(p_{i,t_{1}},p_{s,t_{2}})=\E(p_{i,t_{1}}p_{s,t_{2}})-
        \E(p_{i,t_{1}})\E(p_{s,t_{2}})=
        \left\{
            \begin{array}{lc}
                0, & \mathrm{if}\quad t_{1}\neq t_{2} \\
                \varphi_{i,s}-\phi_{i}\phi_{s}, & \mathrm{if}\quad t_{1}=t_{2},
            \end{array}
        \right.
        \nonumber
    \end{equation}
    where
        \begin{equation}
        \varphi_{i,s}=\int_{\R^{2}}
        \left[\frac{\exp(z_{i})\exp(z_{s})}{\{1+\exp(z_{i})\}\{1+\exp(z_{s})\}}\right]
        t_{2}(\bz_{A}\mid\nu_{0},\bmu_{A,0},\bSigma_{A,0})
        \diff\bz_{A}.
        \nonumber
    \end{equation}
    Finally,
    \begin{equation}
        \Cov(k_{i}-1,k_{s}-1)=\sum_{t=1}^{n-1}\Cov(p_{i,t},p_{s,t})=
        (n-1)(\varphi_{i,s}-\phi_{i}\phi_{s}).
        \nonumber
    \end{equation}
    
    \subsection*{Proof of Proposition 3.} Consider two disjoint non-empty subsets $A,B$ of
    $\{1,\ldots,L\}$ and set $\bc_{D,t}=(c_{i,t}:i\in D)^{\top}$ for any non-empty subset
    $D$ of $\{1,\ldots,L\}$. Then,
    \begin{equation}
        \pi(\bc_{A,t}\mid\bc_{B,t})=\frac{\pi(\bc_{A\cup B,t})}{\pi(\bc_{B,t})}.
        \nonumber
    \end{equation}
    Using the Change of Variables theorem,
    \begin{equation}
        \pi(\bc_{B,t})=\int_{\R^{L}}
        \left\{\prod_{i\in B}\frac{\exp(c_{i,t}z_{i})}{1+\exp(z_{i})}\right\}
        t_{L}(\bz\mid\nu_{0},\bmu_{0},\bSigma_{0})\diff\bz,
        \nonumber
    \end{equation}
    where $\bz=(z_{1},\ldots,z_{L})^{\top}$.  Using Result 1 and the notation
    therein, it follows that
    \begin{equation}
        \pi(\bc_{B,t})=\int_{\R^{|B|}}
        \left\{\prod_{i\in B}\frac{\exp(c_{i,t}z_{i})}{1+\exp(z_{i})}\right\}
        t_{|B|}(\bz_{B}\mid\nu_{0},\bmu_{B,0},\bSigma_{B,0})\diff\bz_{B}.
        \nonumber
    \end{equation}
    Similar expression applies for $\pi(\bc_{A\cup B,t})$. Therefore,
    \begin{equation}
        \pi(\bc_{A,t}\mid\bc_{B,t})=\frac{\int_{\R^{|A|+|B|}}
        \left\{\prod_{i\in A\cup B}\frac{\exp(c_{i,t}z_{i})}{1+\exp(z_{i})}\right\}
        t_{|A|+|B|}(\bz_{A\cup B}\mid\nu_{0},\bmu_{A\cup B,0},\bSigma_{A\cup B,0})
        \diff\bz_{A\cup B}}
        {\int_{\R^{|B|}}
        \left\{\prod_{i\in B}\frac{\exp(c_{i,t}z_{i})}{1+\exp(z_{i})}\right\}
        t_{|B|}(\bz_{B}\mid\nu_{0},\bmu_{B,0},\bSigma_{B,0})\diff\bz_{B}}.
        \nonumber
    \end{equation}
    As for the conditional probability $\Pb(c_{i,t}=1\mid c_{s,t}=1)$, take
    $A=\{i\}$ and $B=\{s\}$.

\bibliographystyle{ASA}
\bibliography{Bibliography}

\begin{thebibliography}{43}
\newcommand{\enquote}[1]{``#1''}
\expandafter\ifx\csname natexlab\endcsname\relax\def\natexlab#1{#1}\fi

\bibitem[{Arellano-Valle et~al.(2013)Arellano-Valle, Castro, and
  Loschi}]{ArellanoValleEtAl2013}
Arellano-Valle, R., Castro, L.~M., and Loschi, R. (2013), \enquote{Change point
  detection in the skew-normal model parameters,} \textit{Communications in
  Statistics. Theory and Methods}, 42, 603--618.

\bibitem[{Barry and Hartigan(1992)}]{Barry1992}
Barry, D. and Hartigan, J.~A. (1992), \enquote{Product partition models for
  change point problems,} \textit{Ann. Statist.}, 20, 260--279.

\bibitem[{Cheon and Kim(2010)}]{Cheon2010}
Cheon, S. and Kim, J. (2010), \enquote{Multiple Change-Point Detection of
  Multivariate Mean Vectors with the Bayesian Approach,} \textit{Comput. Stat.
  Data Anal.}, 54, 406–--415.

\bibitem[{Dahl et~al.(2021{\natexlab{a}})Dahl, Johnson, and
  Mueller}]{dahl2021search}
Dahl, D.~B., Johnson, D.~J., and Mueller, P. (2021{\natexlab{a}}),
  \enquote{Search Algorithms and Loss Functions for Bayesian Clustering,}
  {arXiv}:2105.04451v1.

\bibitem[{Dahl et~al.(2021{\natexlab{b}})Dahl, Johnson, and
  Müller}]{salso:2021}
Dahl, D.~B., Johnson, D.~J., and Müller, P. (2021{\natexlab{b}}),
  \textit{salso: Search Algorithms and Loss Functions for Bayesian Clustering},
  r package version 0.2.23.

\bibitem[{de~P.~Filleti et~al.(2008)de~P.~Filleti, Hotta, and
  Zevallos}]{FilletiEtAl2008}
de~P.~Filleti, J., Hotta, L.~K., and Zevallos, M. (2008), \enquote{Analysis of
  contagion in emerging markets,} \textit{Journal of Data Science}, 6,
  601--626.

\bibitem[{Ding(2016)}]{Ding2016}
Ding, P. (2016), \enquote{On the Conditional Distribution of the Multivariate
  \textit{t} Distribution,} \textit{The American Statistician}, 70, 293--295.

\bibitem[{Erdman and Emerson(2007)}]{bcp_Rpackage}
Erdman, C. and Emerson, J.~W. (2007), \enquote{{bcp}: An {R} Package for
  Performing a Bayesian Analysis of Change Point Problems,} \textit{Journal of
  Statistical Software}, 23, 1--13.

\bibitem[{Fan and Mackey(2017)}]{Fan:2017}
Fan, Z. and Mackey, L. (2017), \enquote{Empirical Bayesian Analysis of
  Simultaneous Changepoints in Multiple Data Sequences,} \textit{Annals of
  Applied Statistics}, 11, 2200--2221.

\bibitem[{Garc\'ia and {Guti\'errez-Pe\~na}(2019)}]{garcia2019}
Garc\'ia, E.~C. and {Guti\'errez-Pe\~na}, E. (2019), \enquote{Nonparametric
  product partition models for multiple change-points analysis,}
  \textit{Communications in Statistics - Simulation and Computation}, 48,
  1922--1947.

\bibitem[{Gupta et~al.(2021)Gupta, Gupta, and Singh}]{Gupta_etal:2021}
Gupta, S.~K., Gupta, N., and Singh, V.~P. (2021), \enquote{Variable-Sized
  Cluster Analysis for 3D Pattern Characterization of Trends in Precipitation
  and Change-Point Detection,} \textit{Journal of Hydrologic Engineering}, 26,
  04020056.

\bibitem[{Harl\'e et~al.(2016)Harl\'e, Chatelain, Gouy-Pailler, and
  Achard}]{harle2016}
Harl\'e, F., Chatelain, F., Gouy-Pailler, C., and Achard, S. (2016),
  \enquote{Bayesian Model for Multiple Change-Points Detection in Multivariate
  Time Series,} \textit{IEEE Transactions on Signal Processing}, 64,
  4351--4362.

\bibitem[{Harrigan(2000)}]{Harrigan2000}
Harrigan, J. (2000), \enquote{The impact of the {A}sia crisis on {U.S}
  industry: {A}n almost-free lunch?} Federal Reserve Bank of New York, Economic
  Policy Review.

\bibitem[{{International Monetary Found}(2003)}]{FMI2003}
{International Monetary Found} (2003), \enquote{El papel del {FMI} en la
  {A}rgentina, 1991-2002,} .

\bibitem[{James et~al.(2019)James, Zhang, and Matteson}]{ecp_Rpackage}
James, N.~A., Zhang, W., and Matteson, D.~S. (2019), \enquote{{ecp}: An {R}
  Package for Nonparametric Multiple Change Point Analysis of Multivariate
  Data. R package version 3.1.2,} .

\bibitem[{Jin et~al.(2021)Jin, Yin, Yuan, and Jiang}]{jin2021}
Jin, H., Yin, G., Yuan, B., and Jiang, F. (2021), \enquote{Bayesian
  Hierarchical Model for Change Point Detection in Multivariate Sequences,}
  \textit{Technometrics}, 0, 1--30.

\bibitem[{Jones et~al.(2021)Jones, Clayton, Ribalet, Armbrust, and
  Harchaoui}]{jones_etal:2021}
Jones, C., Clayton, S., Ribalet, F., Armbrust, E.~V., and Harchaoui, Z. (2021),
  \enquote{A kernel-based change detection method to map shifts in
  phytoplankton communities measured by flow cytometry,} \textit{Methods in
  Ecology and Evolution}, 12, 1687--1698.

\bibitem[{Killick et~al.(2012)Killick, Fearnhead, and Eckley}]{737745}
Killick, R., Fearnhead, P., and Eckley, I.~A. (2012), \enquote{Optimal
  Detection of Changepoints With a Linear Computational Cost,} \textit{Journal
  of the American Statistical Association}, 107, 1590--1598.

\bibitem[{Loschi and Cruz(2002)}]{Loschi2002}
Loschi, R. and Cruz, F. (2002), \enquote{Analysis of the influence of some
  prior specifications in the identification of change points via product
  partition model,} \textit{Computational Statistics \& Data Analysis}, 39,
  477--501.

\bibitem[{Loschi and Cruz(2005)}]{Loschi2005}
--- (2005), \enquote{Extension to the product partition model: computing the
  probability of a change,} \textit{Computational Statistics \& Data Analysis},
  48, 255--268.

\bibitem[{Loschi et~al.(2005)Loschi, Cruz, and Arellano-Valle}]{LoschiCruz2005}
Loschi, R., Cruz, F., and Arellano-Valle, R. (2005), \enquote{Multiple change
  point analysis for the regular exponential family using the product partition
  model,} \textit{Journal of Data Science}, 3, 305--330.

\bibitem[{Loschi et~al.(2003)Loschi, Cruz, Iglesias, and
  Arellano-Valle}]{Loschi2003}
Loschi, R., Cruz, F., Iglesias, P., and Arellano-Valle, R. (2003), \enquote{A
  Gibbs sampling scheme to product partition model: An application to
  change-point problems,} \textit{Computers \& Operations Research}, 30,
  463--482.

\bibitem[{Loschi et~al.(2010)Loschi, Pontel, and Cruz}]{Loschi2010}
Loschi, R., Pontel, J., and Cruz, F. (2010), \enquote{Multiple change-point
  analysis for linear regression models,} \textit{Chilean Journal of
  Statistics}, 1, 93--112.

\bibitem[{Lowell et~al.(1998)Lowell, Neu, and Tong}]{MR-962}
Lowell, J.~F., Neu, C.~R., and Tong, D. (1998), \textit{Financial Crises and
  Contagion in Emerging Market Countries}, Santa Monica, CA: RAND Corporation.

\bibitem[{Mart\'inez and Mena(2014)}]{Martinez2014}
Mart\'inez, A.~F. and Mena, R.~H. (2014), \enquote{On a Nonparametric Change
  Point Detection Model in Markovian Regimes,} \textit{Bayesian Analysis}, 9,
  823--858.

\bibitem[{Matteson and James(2014)}]{matteson2014}
Matteson, D. and James, N. (2014), \enquote{A Nonparametric Approach for
  Multiple Change Point Analysis of Multivariate Data,} \textit{Journal of the
  American Statistical Association}, 109, 334--345.

\bibitem[{Meil\v{a}(2007)}]{VI2007}
Meil\v{a}, M. (2007), \enquote{Comparing clusterings-an information based
  distance,} \textit{Journal of Multivariate Analysis}, 98, 873--895.

\bibitem[{Nyamundanda et~al.(2015)Nyamundanda, Hegarty, and
  Hayes}]{Nyamundanda2015}
Nyamundanda, G., Hegarty, A., and Hayes, K. (2015), \enquote{Product partition
  latent variable model for multiple change-point detection in multivariate
  data,} \textit{Journal of Applied Statistics}, 42, 2321--2334.

\bibitem[{Padilla et~al.(2021)Padilla, Yu, Wang, and Rinaldo}]{EJS1809}
Padilla, O. H.~M., Yu, Y., Wang, D., and Rinaldo, A. (2021), \enquote{{Optimal
  nonparametric change point analysis},} \textit{Electronic Journal of
  Statistics}, 15, 1154--1201.

\bibitem[{Page and Quinlan(2022)}]{ppmSuite}
Page, G.~L. and Quinlan, J.~J. (2022), \textit{ppmSuite: A Collection of Models
  that Employ a Product Parition Prior Distribution on Partitions}, r package
  version 0.2.1.

\bibitem[{Page et~al.(2021)Page, Quintana, and Dahl}]{page2021}
Page, G.~L., Quintana, F.~A., and Dahl, D.~B. (2021), \enquote{Dependent
  Modeling of Temporal Sequences of Random Partitions,} \textit{Journal of
  Computational and Graphical Statistics}, 0, 1--29.

\bibitem[{Pedroso et~al.(2021)Pedroso, Loschi, and
  Quintana}]{pedroso2021multipartition}
Pedroso, R.~C., Loschi, R.~H., and Quintana, F.~A. (2021),
  \enquote{Multipartition model for multiple change point identification,}
  {arXiv}:2107.11456v2.

\bibitem[{{R Core Team}(2021)}]{R}
{R Core Team} (2021), \textit{R: A Language and Environment for Statistical
  Computing}, R Foundation for Statistical Computing, Vienna, Austria.

\bibitem[{Rand(1971)}]{ARI}
Rand, W.~M. (1971), \enquote{Objective Criteria for the Evaluation of
  Clustering Methods,} \textit{Journal of the American Statistical
  Association}, 66, 846--850.

\bibitem[{Stallings(1998)}]{Stalligns1998}
Stallings, B. (1998), \enquote{Impact of th {A}sian crisis on {L}atin
  {A}merica,} United Nations Economic Commission for Latin America and the
  Caribbean.

\bibitem[{Truong et~al.(2020)Truong, Oudre, and Vayatis}]{truong:2020}
Truong, C., Oudre, L., and Vayatis, N. (2020), \enquote{Selective review of
  offline change point detection methods,} \textit{Signal Processing}, 167,
  107299.

\bibitem[{Tveten et~al.(2021)Tveten, Eckley, and
  Fearnhead}]{tveten2021scalable}
Tveten, M., Eckley, I.~A., and Fearnhead, P. (2021), \enquote{Scalable
  changepoint and anomaly detection in cross-correlated data with an
  application to condition monitoring,} {arXiv}:2010.06937v2.

\bibitem[{Vald\'es(2000)}]{Valdes2000}
Vald\'es, R. (2000), \enquote{Emerging markets contagion: evidence and theory,}
  Available at SSRN: https://ssrn.com/abstract=69093 or
  http://dx.doi.org/10.2139/ssrn.69093.

\bibitem[{Wang and Emerson(2015)}]{wang2015bayesian}
Wang, X. and Emerson, J.~W. (2015), \enquote{Bayesian Change Point Analysis of
  Linear Models on Graphs,} {a}rXiv:1509.00817.

\bibitem[{Wang(1993)}]{Wang1993}
Wang, Y.~H. (1993), \enquote{On the Number of Successes in Independent Trials,}
  \textit{Statistica Sinica}, 3, 295--312.

\bibitem[{Wood et~al.(2021)Wood, Roberts, and Zohren}]{2021}
Wood, K., Roberts, S., and Zohren, S. (2021), \enquote{Slow Momentum with Fast
  Reversion: A Trading Strategy Using Deep Learning and Changepoint Detection,}
  \textit{The Journal of Financial Data Science}, jfds.2021.1.081.

\bibitem[{Yao(1984)}]{Yao1984}
Yao, Y.-C. (1984), \enquote{Estimation of a noisy discrete-time step function:
  Bayes and empirical Bayes approaches,} \textit{Ann. Statist.}, 12,
  1434--1447.

\bibitem[{Zanini et~al.(2019)Zanini, M{\"u}ller, Ji, and Quintana}]{carlosetal}
Zanini, C. T.~P., M{\"u}ller, P., Ji, Y., and Quintana, F.~A. (2019),
  \enquote{A Bayesian Random Partition Model for Sequential Refinement and
  Coagulation,} \textit{Biometrics}, 75, 988--999.

\end{thebibliography}

\end{document}